\PassOptionsToPackage{usenames,dvipsnames}{xcolour}
\documentclass[twocolumn,twocolappendix,nofootinbib,iop]{openjournal}

\pdfoutput=1 %for arXiv submission
\usepackage{amsmath,amssymb,amstext}
\usepackage[T1]{fontenc}
\usepackage{apjfonts}
\usepackage{ae,aecompl}
\usepackage[utf8]{inputenc}
\usepackage[colorlinks,allcolors=blue]{hyperref}
\usepackage[figure,figure*]{hypcap}
\usepackage{natbib}
\usepackage{url}
\usepackage{mdwlist}
\usepackage{multirow}
\usepackage{lineno}
\usepackage{fontawesome}

\usepackage{soul}

\usepackage{xcolor}
\definecolor{maroon}{cmyk}{0, 0.87, 0.68, 0.32}
\definecolor{halfgray}{gray}{0.55}
\definecolor{ipython_frame}{RGB}{207, 207, 207}
\definecolor{ipython_bg}{RGB}{247, 247, 247}
\definecolor{ipython_red}{RGB}{186, 33, 33}
\definecolor{ipython_green}{RGB}{0, 128, 0}
\definecolor{ipython_cyan}{RGB}{64, 128, 128}
\definecolor{ipython_purple}{RGB}{170, 34, 255}

\usepackage{listings}

\lstdefinelanguage{iPython}{
    morekeywords={access,and,break,class,continue,def,del,elif,else,except,exec,finally,for,from,global,if,import,in,is,lambda,not,or,pass,print,raise,return,try,while},%
    morekeywords=[2]{abs,all,any,basestring,bin,bool,bytearray,callable,chr,classmethod,cmp,compile,complex,delattr,dict,dir,divmod,enumerate,eval,execfile,file,filter,float,format,frozenset,getattr,globals,hasattr,hash,help,hex,id,input,int,isinstance,issubclass,iter,len,list,locals,long,map,max,memoryview,min,next,object,oct,open,ord,pow,property,range,raw_input,reduce,reload,repr,reversed,round,set,setattr,slice,sorted,staticmethod,str,sum,super,tuple,type,unichr,unicode,vars,xrange,zip,apply,buffer,coerce,intern},%
    sensitive=true,%
    morecomment=[l]\#,%
    morestring=[b]',%
    morestring=[b]",%
    morestring=[s]{'''}{'''},% used for documentation text (mulitiline strings)
    morestring=[s]{"""}{"""},% added by Philipp Matthias Hahn
    morestring=[s]{r'}{'},% `raw' strings
    morestring=[s]{r"}{"},%
    morestring=[s]{r'''}{'''},%
    morestring=[s]{r"""}{"""},%
    morestring=[s]{u'}{'},% unicode strings
    morestring=[s]{u"}{"},%
    morestring=[s]{u'''}{'''},%
    morestring=[s]{u"""}{"""},%
    literate=
    {á}{{\'a}}1 {é}{{\'e}}1 {í}{{\'i}}1 {ó}{{\'o}}1 {ú}{{\'u}}1
    {Á}{{\'A}}1 {É}{{\'E}}1 {Í}{{\'I}}1 {Ó}{{\'O}}1 {Ú}{{\'U}}1
    {à}{{\`a}}1 {è}{{\`e}}1 {ì}{{\`i}}1 {ò}{{\`o}}1 {ù}{{\`u}}1
    {À}{{\`A}}1 {È}{{\'E}}1 {Ì}{{\`I}}1 {Ò}{{\`O}}1 {Ù}{{\`U}}1
    {ä}{{\"a}}1 {ë}{{\"e}}1 {ï}{{\"i}}1 {ö}{{\"o}}1 {ü}{{\"u}}1
    {Ä}{{\"A}}1 {Ë}{{\"E}}1 {Ï}{{\"I}}1 {Ö}{{\"O}}1 {Ü}{{\"U}}1
    {â}{{\^a}}1 {ê}{{\^e}}1 {î}{{\^i}}1 {ô}{{\^o}}1 {û}{{\^u}}1
    {Â}{{\^A}}1 {Ê}{{\^E}}1 {Î}{{\^I}}1 {Ô}{{\^O}}1 {Û}{{\^U}}1
    {œ}{{\oe}}1 {Œ}{{\OE}}1 {æ}{{\ae}}1 {Æ}{{\AE}}1 {ß}{{\ss}}1
    {ç}{{\c c}}1 {Ç}{{\c C}}1 {ø}{{\o}}1 {å}{{\r a}}1 {Å}{{\r A}}1
    {€}{{\EUR}}1 {£}{{\pounds}}1
    {^}{{{\color{ipython_purple}\^{}}}}1
    {=}{{{\color{ipython_purple}=}}}1
    {+}{{{\color{ipython_purple}+}}}1
    {-}{{{\color{ipython_purple}-}}}1
    {*}{{{\color{ipython_purple}$^\ast$}}}1
    {/}{{{\color{ipython_purple}/}}}1
    {+=}{{{+=}}}1
    {-=}{{{-=}}}1
    {*=}{{{$^\ast$=}}}1
    {/=}{{{/=}}}1,
    literate=
    *{-}{{{\color{ipython_purple}-}}}1
     {?}{{{\color{ipython_purple}?}}}1,
    identifierstyle=\color{black}\ttfamily,
    commentstyle=\color{ipython_cyan}\ttfamily,
    stringstyle=\color{ipython_red}\ttfamily,
    keepspaces=true,
    showspaces=false,
    showstringspaces=false,
    rulecolor=\color{ipython_frame},
    frameround={t}{t}{t}{t},
    numbers=none,
    numberstyle=\tiny\color{halfgray},
    backgroundcolor=\color{ipython_bg},
    basicstyle=\ttfamily\footnotesize,
    columns=fullflexible,
    keywordstyle=\color{ipython_green}\ttfamily,
}

\newcommand{\nblink}[1]{\href{https://github.com/DifferentiableUniverseInitiative/jax-cosmo-paper/blob/master/notebooks/#1.ipynb}{\faFileCodeO}}
\newcommand{\github}{\href{https://github.com/DifferentiableUniverseInitiative/jax\_cosmo}{\faGithub}}

\newcommand{\resubnote}[1]{#1} %JEC after paper was accept 27 April 2023.

\newcommand{\numpyro}{\texttt{NumPyro}}
\newcommand{\bydef}{:=}
\newcommand{\jaxcosmo}{\texttt{jax-cosmo}}
\newcommand{\autodiff}{\texttt{autodiff}}
\newcommand{\jax}{\texttt{JAX}}

\begin{document}
\journalinfo{The Open Journal of Astrophysics}
%\submitted{submitted xxx, revised yyy, accepted zzz}

\title{\texttt{JAX-COSMO}: An End-to-End Differentiable and GPU Accelerated Cosmology Library}


\author{
J.~E. Campagne$^{1,\ast}$, 
F. Lanusse$^2$,
J. Zuntz$^3$,\\
A. Boucaud$^4$,
S. Casas$^{5}$,
M.~Karamanis$^{6,7}$,
D.~Kirkby$^8$,
D. Lanzieri$^9$,
Y.~Li$^{10,11}$,
A. Peel$^{12}$
}
\thanks{$^\ast$jean-eric.campagne@ijclab.in2p3.fr}

\affiliation{
$^1$Université Paris-Saclay, CNRS/IN2P3, IJCLab, 91405 Orsay, France
}
\affiliation{$^2$Université Paris-Saclay, Université Paris Cité, CEA, CNRS, AIM, 91191, Gif-sur-Yvette, France}
\affiliation{$^3$Institute for Astronomy, University of Edinburgh, Edinburgh EH9 3HJ, United Kingdom}
\affiliation{$^4$Université de Paris, CNRS, Astroparticule et Cosmologie, F-75013 Paris, France}
\affiliation{$^{5}$Institute for Theoretical Particle Physics and Cosmology (TTK), RWTH Aachen University, 52056 Aachen, Germany.}
\affiliation{$^{6}$Berkeley Center for Cosmological Physics, University of California, Berkeley, CA 94720, USA}
\affiliation{$^{7}$Lawrence Berkeley National Laboratory, 1 Cyclotron Road, Berkeley, CA 94720, USA}
\affiliation{$^8$Department of Physics and Astronomy, University of California, Irvine, CA 92697, USA}
\affiliation{$^9$Université Paris Cité, Université Paris-Saclay, CEA, CNRS, AIM, F-91191, Gif-sur-Yvette, France}
\affiliation{$^{10}$Department of Mathematics and Theory, Peng Cheng Laboratory, Shenzhen, Guangdong 518066, China}
\affiliation{$^{11}$Center for Computational Astrophysics \& Center for Computational Mathematics, Flatiron Institute, New York, New York 10010, USA}
\affiliation{$^{12}$Ecole Polytechnique F\'ed\'erale de Lausanne (EPFL), Observatoire de Sauverny, 1290 Versoix, Switzerland}

%\date{\today}

\begin{abstract}
We present \jaxcosmo, a library for automatically differentiable cosmological theory calculations. \jaxcosmo\ uses the \jax\ library, which has created a new coding ecosystem, especially in probabilistic programming.
As well as batch acceleration, just-in-time compilation, and automatic optimization of code for different hardware modalities (CPU, GPU, TPU), \jax\ exposes an \textit{automatic differentiation} (autodiff) mechanism. Thanks to autodiff, \jaxcosmo\ gives access to the derivatives of cosmological likelihoods with respect to any of their parameters, and thus enables a range of powerful %  AD can take a Python function that uses \jax\ and return its (vector) derivative, computed not with finite differences but by successively differentiating each instruction in turn. This gives us the opportunity to apply a range of powerful
Bayesian inference algorithms, otherwise impractical in cosmology, such as Hamiltonian Monte Carlo and Variational Inference. 
In its initial release, \jaxcosmo\ implements background evolution, linear and non-linear power spectra (using \texttt{halofit} or the Eisenstein and Hu transfer function), as well as angular power spectra ($C_\ell$) with the Limber approximation for galaxy and weak lensing probes, all differentiable with respect to the cosmological parameters and their other inputs. 
We illustrate how automatic differentiation can be a game-changer for common tasks involving Fisher matrix computations, or full posterior inference with gradient-based techniques (e.g. Hamiltonian Monte Carlo). In particular, we show how Fisher matrices are now fast, exact, no longer require any fine tuning, and are themselves differentiable with respect to parameters of the likelihood, enabling complex survey optimization by simple gradient descent. Finally, using a Dark Energy Survey Year 1 3x2pt analysis as a benchmark, we demonstrate how \jaxcosmo\ can be combined with Probabilistic Programming Languages such as \numpyro\ to perform posterior inference with state-of-the-art algorithms including a No U-Turn Sampler (NUTS), Automatic Differentiation Variational Inference (ADVI), and Neural Transport HMC (NeuTra). 
% We discuss algorithms made possible by this library, present comparisons with the Core Cosmology Library as a benchmark, and run a series of tests using the Dark Energy Survey Year 1 3x2pt analysis with the \numpyro\ library to demonstrate practical inference. 
We show that the effective sample size per node (1 GPU or 32 CPUs) per hour of wall time is about 5 times better for a JAX NUTS sampler compared to the well optimized Cobaya Metropolis-Hasting sampler. We further demonstrate that Normalizing Flows using Neural Transport are a promising methodology for model validation in the early stages of analysis.   %
\github

% The recent \jax\ library has created a new ecosystem from which probabilistic programming software can take enormous benefit: batch acceleration, just-in-time compilation, and automatic optimization of code for different hardware modalities (CPU, GPU, TPU) can provide huge speed-ups to a wide range of different problems. In particular, \jax\ exposes an \textit{automatic differentiation} mechanism, which can take a python function that uses \jax\ and return its (vector) derivative.  This gives us the opportunity to apply a range of powerful but otherwise unfeasible algorithms used in Bayesian inference, such as Hamiltonian Monte Carlo (HMC) and Variational Inference. 

% To take advantage of these possibilities within cosmology we have developed the \jaxcosmo\ library, which implements background evolution, linear and non linear power spectra (using \texttt{halofit} or the Eisenstein and Hu transfer function), as well as angular power spectra ($C_\ell$) with the Limber approximation for galaxy and weak lensing probes. We discuss algorithms made possible by this library, and present comparisons with the Core Cosmology Library as a benchmark, and run a series of tests using the Dark Energy Survey Year 1 3x2pt analysis with the \numpyro\ library to demonstrate practical usage. 
% We show that clear improvements are possible using HMC compared to Metropolis-Hasting, and that the Normalizing Flows using the Neural Transport is a promising methodology.
%{\color{red}
%;in the early phases of Bayesian model development JAZ not sure what this means so I left it out}

\end{abstract}

\maketitle



\section{Introduction}
Bayesian inference has been widely used in cosmology in the form of Monte Carlo Markov Chains (MCMC) since the work of \citet{2001ApJ...563L..95K} and \citet{2003MNRAS.341.1084R}, and has been the keystone for past and current analysis thanks partly to packages such as  \texttt{CosmoMC} \citep{2002PhRvD..66j3511L}, \texttt{CosmoSIS} \citep{2015A&C....12...45Z}, \texttt{MontePython} \citep{2019PDU....24..260B}, and \texttt{Cobaya} \citep{2019ascl.soft10019T,2021JCAP...05..057T}; see, for instance, the list of citations to these popular packages for an idea of the wide usage in the community. 

% \footnote{\url{https://cosmologist.info/cosmomc/readme.html}}
% \footnote{\url{http://bitbucket.org/joezuntz/cosmosis}}
% \footnote{\url{https://cobaya.readthedocs.io}}
% \footnote{\url{https://github.com/brinckmann/montepython_public}}

% Note that some direct optimization of likelihood function has also been successfully undertaken for instance for some Planck analysis \citep{2014A&A...566A..54P}, 
Since the development of these MCMC packages, major advances have been made in automatic differentiation (\textit{autodiff}) \citep{10.5555/3122009.3242010, Margossian2019}, a set of technologies for transforming pieces of code into their derivatives.

While these tools have especially been applied to neural network optimization and machine learning (ML) in general, they can also enable classical statistical methods that require the derivatives of (e.g. likelihood) functions to operate: we consider such methods in this paper. \textit{Autodiff} has been implemented in widely used libraries like \texttt{Stan} \citep{JSSv076i01}, \texttt{TensorFlow} \citep{tensorflow2015-whitepaper}, \texttt{Julia} \citep{bezanson2017julia}, and \texttt{PyTorch} \citep{NEURIPS2019_9015}. 

A recent entrant to this field is the \jax\ library\footnote{\url{https://jax.readthedocs.io}} \citep{jax2018github} which has undergone rapid development and can automatically differentiate native \texttt{Python} and \texttt{NumPy} functions, offering a speed up to the development process and indeed code runtimes. \jax\ offers an easy parallelization mechanism (\texttt{vmap}), just-in-time compilation (\texttt{jit}), and optimization targeting CPU, GPU, and TPU hardware thanks to the \texttt{XLA} library (which also powers TensorFlow). These attractive features have driven wide adoption of \jax\ in computational research, and motivate us to consider its usage in cosmology.

\jax\ contains bespoke reimplementations of packages such as \texttt{jax.numpy} and \texttt{jax.scipy}, as well as example libraries such as \texttt{Stax} for simple but flexible neural network development. It has formed the seed for a wider ecosystem of packages, including, for example: 
\texttt{Flax} \citep{flax2020github} a high-performance neural network library, \texttt{JAXopt} \citep{jaxopt_implicit_diff} a hardware accelerated, batchable and differentiable collection of optimizers, \texttt{Optax} \citep{optax2020github} a gradient processing and optimization library, and \numpyro\ \citep{phan2019composable,bingham2019pyro}, a probabilistic programming language (PPL) that is used in this paper. Other PPL packages such as \texttt{PyMC} \citep{Salvatier2016} have switched to a \jax\ backend in recent versions\footnote{A more exhaustive list and rapidly growing list of packages can be found at \url{https://project-awesome.org/n2cholas/awesome-jax}}.

% I don't think we need all these URLs, it's getting excessive, especially when they have papers:
%\footnote{\url{https://flax.readthedocs.io/}}
% \footnote{\url{https://jaxopt.github.io}}
% \footnote{\url{https://optax.readthedocs.io}}
% \footnote{\url{https://num.pyro.ai}}
% \footnote{\url{https://www.pymc.io/projects/docs/en/stable/installation.html}}
% \footnote{\url{https://www.tensorflow.org/xla}}
% \footnote{\url{https://www.tensorflow.org/}}
% \footnote{\url{https://pytorch.org/docs/}}
% \footnote{\url{https://mc-stan.org/users/documentation}}

To explore alternative inference methods to the well-known Metropolis-Hasting likelihood sampler, and in order to use GPU devices in the context of JAX framework, we have developed the open source \jaxcosmo\ library\footnote{\url{https://github.com/DifferentiableUniverseInitiative/jax_cosmo}}, which we present in this paper. The package represents a first step in making the powerful features described above useful for cosmology; it implements a selection of theory predictions for key cosmology observables as differentiable \jax\ functions.

%\FrL{Missing a paragraph on what we propose to do with this framework in this paper, and why that's interesting in the context of the very long time that it takes to sample common likelihood.} 

We give an overview of the code's design and contents in Section~\ref{sec-jaxcosmo-design}. We show how to use it for rapid and numerically stable Fisher forecasts in Section~\ref{sec-fisher-forecast}, in more complete gradient-based cosmological inference with variants of Hamiltonian Monte Carlo including the No-U-Turn Sampler, and ML-accelerated Stochastic Variational Inference in Section~\ref{sec:chmc}. We discuss and compare these methods in Section~\ref{sec-discussion} and conclude in Section~\ref{sec-conclusion}.

\section{\jax: GPU Accelerated and Automatically Differentiable Python Programming}
\label{sec-primer}

The aim of this section is to provide a brief technical primer on \jax, necessary to fully grasp the potential of a cosmology library implemented in this framework.

% Maybe a short history explaining what autograd and XLA are.
%\FrL{Add a short history of XLA and Autograd.} Or maybe not

\paragraph{\textbf{Automatic Differentiation}} Traditionally, two different approaches have been used in cosmology to obtain derivatives of given numerical expressions. The first approach is to derive analytically the formula for the derivatives of interest \citep[e.g.][]{2013MNRAS.432..894J, kostic}, with or without the help of tools such as \texttt{Mathematica}\footnote{\url{https://www.wolfram.com/mathematica}.}. This is typically only practical, however, for simple analytical models.
The second approach is to compute numerical derivatives by finite differences. This approach can be applied on black-box models of arbitrary complexity (from typical Boltzmann codes to cosmological simulations; \citealp{2020ApJS..250....2V}). However it is notoriously difficult to obtain stable derivatives by finite differences \citep[e.g.][]{2021arXiv210100298B, 2021A&A...649A..52Y}. In addition, their computational cost does not scale well with the number of parameters (a minimum of $(2N+1)$ model evaluations is typically required for $N$ parameters), making them impractical whenever derivatives are needed as part of an outer iterative algorithm.

Automatic differentiation frameworks like \jax\ take a different approach. They trace the execution of a given model and decompose this trace into primitive operations with known derivatives (e.g. multiplication). Then, by applying the chain rule formula, the computational graph for the derivatives (of any order) of the model can be built from the known derivatives of every elementary operation. A new function corresponding to the derivative of the original function is therefore built automatically for the user. We direct the interested reader to \citet{baydin2018automatic} and \citet{Margossian2019} for in-depth introductions to automatic differentiation. 

\jax\ provides in particular a number of operators (\texttt{jax.grad}, \texttt{jax.jacobian}, \texttt{jax.hessian}) which can compute derivatives of any function written in \jax:
\begin{lstlisting}[language=iPython]
# Define a simple function 
def f(x):
	return y = 5 * x + 2
# Take the derivative
df_dx = jax.grad(f)
# df_dx is a new function that always returns 5 
\end{lstlisting}

\textbf{Why is this interesting?} \autodiff\ makes it possible to obtain \textit{exact gradients of cosmological likelihoods} with respect to all input parameters at the cost of only two likelihood evaluations.

\paragraph{\textbf{Just In Time Compilation (JIT)}} Despite its convenience and wide adoption in astrophysics, Python still suffers from slow execution times compared to fully compiled languages such as C/C++. One approach to mitigate these issues and make Python code fast is Just In Time compilation, which traces the execution of a given Python function the first time it is called, and compiles it into a fast executable (by-passing the Python interpreter), which can be transparently used in subsequent calls to this function. Compared to other strategies for speeding up Python code such as Cython, JIT allows the user to write plain Python code, and reap the benefits of compiled code.

A number of libraries allowing for JIT have already been used in astrophysics, in particular Numba\footnote{\url{https://numba.pydata.org/}}, or the HOPE library \cite{2015A&C....10....1A} developed specifically for the needs of astrophysics. \jax\ stands out compared to these other frameworks in that it relies on the highly optimized XLA library\footnote{\url{https://www.tensorflow.org/xla}} for executing the compiled expressions. XLA is continuously developed by Google as part of the TensorFlow project, for efficient training and inference of large scale deep learning applications, and as such supports computational backends such as GPU and Tensor Processing Units (TPU) clusters. The ability to perform computations directly on GPUs through XLA is one of the major benefits of \jax, as speed-ups of at least two orders of magnitudes can be expected for typical parallel linear algebra computations compared to CPU.

In \jax, jitting is achieved by transforming a function with the \texttt{jax.jit} operation:
\begin{lstlisting}[language=iPython]
# Redefine our function
def f(x):
	return y = 5 * x + 2
# And JIT it
jitted_f = jax.jit(f)
# The first call to jitted_f will be relatively slow
# subsequent calls will be extremely fast
# and run as a compiled code directly on GPU
\end{lstlisting}

\textbf{Why is this interesting?} JIT makes it possible to execute entire cosmological \textit{MCMC chains directly on GPUs as compiled code}, with orders of magnitude gain in speedup over Python code.

\paragraph{\textbf{Automatic Vectorization}} Another extremely powerful feature of \jax\ is its ability to automatically vectorize or paralellize any function. Through the same tracing mechanism used for automatic differentiation, \jax\ can decompose a given computation into primitive operations and add a new \textit{batch} dimension so that the computation can be applied to a batch of inputs as opposed to individual ones. In doing so, the computation will not run sequentially over all entries of the batch, but truly in parallel making full use of the intrinsic parallel architecture of modern GPUs.

In \jax\ automatic vectorization is achieved using the \textit{jax.vmap} operation:
\begin{lstlisting}[language=iPython]
# Our function f only applies to scalars
def f(x):
	return y = 5 * x + 2
# Applying automatic vectorization
batched_f = jax.vmap(f)
# batched_f now applies to 1D arrays
\end{lstlisting}
Again, we stress that in this example, \texttt{batched\_f} will not be implemented in terms of a for loop, but with operations over vectors. The function above is trivial, but the same approach can be used to parallelize any function, from Limber integrals, to an entire likelihood. For multi-device use-cases (e.g., several GPUs or TPUs), \jax\ provides \texttt{pmap} which compiles and executes, in parallel, replicas of the same code on each device. Moreover, recent experimental developments deal with parallelization of functions over supercomputer-scale hardware meshes. In the examples detailed in this article, we have only relied on \texttt{vmap} functionality.

\textbf{Why is this interesting?} Automatic Vectorization makes it possible to trivially parallelize cosmological likelihood evaluations, to run many parallel MCMC chains on a single GPU.

\paragraph{\textbf{NumPy API compliance}} Finally, the last point to note about \jax, is that it mirrors the NumPy API (with only a few exceptions). This means in practice that existing NumPy code can easily be converted to \jax. This is in contrast to other similar frameworks like TensorFlow, PyTorch, or Julia which all require the user to learn, and adapt their code to, a new API or even a new language. 

\textbf{Why is this interesting?} NumPy compliance implies improved maintainability and lower barrier to entry for new contributors.

\section{Capabilities of the \jaxcosmo\ library}
\label{sec-jaxcosmo-design}

In this section, we describe the cosmological modeling provided by \jaxcosmo, and its implementation in \jax. The general design follows that of the Core Cosmology Library (CCL; \citealp{2019ApJS..242....2C}), though in its initial release
 \jaxcosmo\ only implements a subset of CCL features and options.

All \jaxcosmo\ data structures are organized as \jax\ container objects, which means that two key \jax\ features are automatically available to them: \textit{autodiff} and \textit{vmap}. The vmap feature enables any operation defined in \jax\ (including complicated composite operations) to be applied efficiently as a vector operation. The autodiff feature further makes it possible to take the derivative of any operation, automatically transforming a function that takes $n$ inputs and produces $m$ outputs into a new function that generates an $m \times n$ matrix of partial derivatives.

\jaxcosmo\ implements, for example, Runge-Kutta solvers for ODEs, as well as Simpson and Romberg integration routines through this framework, so that we can automatically compute the derivatives of their solutions with respect to their inputs. This includes not only the cosmological parameters (the standard set of $w_0 w_a CDM$ cosmological parameters is exposed, using $\sigma_8$ as an amplitude parameter), but also other input quantities such as redshift distributions as described below.

In the rest of this section we describe the cosmological calculations that are implemented using these facilities.
\subsection{Formalism \& Implementation}

\subsubsection{Background cosmology}

The computation of the evolution of the cosmological background follows a typical implementation of a Friedmann equation (\citealp[see e.g.][]{2005A&A...443..819P}):
\begin{equation}
    E^2(a) = \frac{\mathrm{H}^2(a)}{\mathrm{H}^2_0} = \Omega_m a^{-3} + \Omega_k a^{-2} + \Omega_{de} e^{f(a)}
\end{equation}
with $a=1/(1+z)$ the scale factor related to the redshift $z$, $\mathrm{H}(a)=\dot{a}/a$ the Hubble parameter with $\mathrm{H}_0$ its present day value, $\Omega_m= \Omega_{cdm}+\Omega_b$, $\Omega_{de}=1-\Omega_k-\Omega_m$, and 
\begin{equation}
    f(a) = -3 (1 + w_0 + w_a) \ln(a) + 3 w_a (a - 1)
\end{equation}
Notably, the relativistic contributions of massless neutrinos and photon radiation, as well as the massive neutrino contribution are currently neglected. From these expressions, in \texttt{jc.background}, are computed the different cosmological distance functions, such as the radial comoving distance
\begin{equation}
     \chi(a) =  R_H \int_a^1 \frac{\mathrm{d}a^\prime}{{a^\prime}^2 E(a^\prime)}
     \label{eq:radial_comoving}
\end{equation}
with $R_H$ the Hubble radius.  \resubnote{$\chi(a)$ evaluation is performed with a linear interpolation of pre-computed values obtained over a logarithmic grid of 256 knots regularly spaced in the range [-3,0] using a Runge-Kutta 4th order solver.}

%Using the scale factor ($a$) redshift ($z$) relationship, $\chi$ can be viewed as a function of $z$. 
%
\subsubsection{Growth of perturbations}
\resubnote{Accurate calculations of the growth of cosmic structure perturbations require solving the Boltzmann equation describing interactions between different density species. This is a complicated process to achieve at high accuracy, and is implemented in dedicated codes such as CAMB \citep{camb} or CLASS \citep{2011JCAP...07..034B}. At this stage we have not attempted to implement a Boltzmann solver in JAX, but instead use the analytic approximation presented in \citet{Eisenstein_1998}. This defines a transfer function $T$} which relates the primordial matter power spectrum to its late-time non-linear value:
\begin{equation}
    P(k, z) = P(k, z=\infty) \cdot T^2(k, z; \Omega_m, \Omega_b, ...),
\end{equation}
through the \textit{halofit} model by \cite{2012ApJ...761..152T} or \cite{2003MNRAS.341.1311S} without the neutrino contribution introduced by \cite{10.1111/j.1365-2966.2011.20222.x}. No Baryon feedback modeling is considered yet. 

The primordial power spectrum is modelled with the standard form:
\begin{equation}
    P(k) = A k^{n_s}.
\end{equation}

The normalisation $A$ is parameterised via $\sigma_8$ at $z=0$ as 
\begin{equation}
    A = \sigma_8^2 \times \left(\frac{1}{2 \pi^2} \int_0^\infty \frac{\mathrm{d}k}{k} k^3 P(k) W^2(kR_8) \right)^{-1}
\end{equation}
with $R_8 = 8 \mathrm{Mpc}/h$ and $W(x)$ related to the $j_1$ spherical Bessel function as
\begin{equation}
    W(x) = \frac{3j_1(x)}{x}
\end{equation}

%Future version of the library would offer the possibility to call a \jax\ emulator of the Cosmic Linear Anisotropy Solving System \citep{2011JCAP...07..034B}. \JZ{How close is this to being usable? If it's just an idea or an early prototype then skip this paragraph.}
%
\subsubsection{Angular power spectra}

\resubnote{The bias and number density functions that enter the kernel terms have been modelled in many different ways to try to capture possible behavior of these key systematic errors. Exposing them in JAX-Cosmo allows us easily to deal with highly complicated models for either of them, opening up many useful avenues of exploration, as shown in \citet{ruiz-zapatero}}.

\jaxcosmo\ is currently focused on predicting projected 2D Fourier-space 2pt galaxy lensing, clustering, and cross correlations, the $C_\ell$ angular power spectra that are a primary target of upcoming photometric surveys. The details of the implementation is in \texttt{jc.angular\_cl} which deals with the mean and Gaussian covariance matrix computations.

The angular power spectra $C_\ell^{ij}$ for the probes $(i,j)$ and for redshift bin window selections are computed in the first order Limber approximation \citep{PhysRevD.78.123506}:
\begin{align}
    C_\ell^{i,j} \approx \left(\ell+\frac{1}{2}\right)^{m_i+m_j}\int_{a_{i,j}}^1\frac{\mathrm{d}a}{c^2\chi^2}\frac{\mathrm{d}\chi}{\mathrm{d}a}K_i(\chi)K_j(\chi)\,P\left(k=\frac{\ell+1/2}{\chi},z\right),\label{eq:Cell_limber}
\end{align}
The $m_i$ factors are $(0,-2)$ for the galaxy clustering and weak lensing, respectively, and each $K(z)$ function represents a single tomographic redshift bin's number density \resubnote{in $z=(1/a)-1$. The integration lower bound $a_{ij}$ is defined from the maximum redshift involved  and  Simpson's rule is used  with a regular grid of 512 knots. The $K(z)$} tracers are implemented as two kernel functions:

\begin{description}
\item[\texttt{NumberCounts}]
    \begin{equation}
        K_i(z) = n_i(z)\ b(z)\ H(z)
    \end{equation}
where $n_i(s)$ is the redshift distribution of the sample (e.g., \texttt{jc.redshift.kde\_nz} function), and $b(z)$ is the galaxy bias function (see  \texttt{jc.bias.constant\_linear\_bias}). No redshift space distortions are taken into account.

\item [\texttt{WeakLensing}]
    \begin{multline}
 K_i(z) = \left( \frac{3 H_0^2\Omega_m}{2 c} \right) \left(\frac{(\ell+2)!}{(\ell-2)!} \right)^{1/2}\times 
 \\  (1+z)\ \chi(z) \int_z^\infty p_i(z^\prime)\ \frac{\chi(z^\prime)-\chi(z)}{\chi(z^\prime)}\ \mathrm{d}z^\prime + K_{IA}(z)
    \end{multline}
where $K_{IA}(z)$ is an optional kernel function to deal with the Intrinsic Alignment. The implementation of this term currently follows \citet{2011A&A...527A..26J}, and is given by:
\begin{equation}
    K_{IA}(z) = \left(\frac{(\ell+2)!}{(\ell-2)!}\right)^{1/2}\ p_i(z)\ b(z)\  H(z)\ \frac{C\  \Omega_m}{D(z)}
\end{equation}
with $C\approx 0.0134$ being a dimensionless constant and $D(z)$ the growth factor of linear perturbations. \resubnote{The integration is performed using Simpson's rule with an upper bound given by the maximum redshift involved.}
\end{description}

% Notice that all described kernel functions can be also viewed as function of the scale factor $a$ or as function of the radial comoving distance $\chi$. 

% \FrL{We probably actually want to start by the Limber formula, which explicits what a kernel is, otherwise people don't necessarily know where these kernels come from.}

Because, like the other ingredients, the \texttt{NumberCounts} and \texttt{WeakLensing} kernels are implemented as JAX objects, all the integrals involved in these computations can be differentiated with respect to the cosmological parameters and the number densities, using \textit{autodiff} and \jaxcosmo's implementation of integration quadrature. An example is given in the context of DES Y1 3x2pts analysis (Sec~\ref{sec-DESY1}).
\subsection{Validation against the Core Cosmology Library (CCL)}
\begin{figure}
    \centering
    \includegraphics[width=\columnwidth]{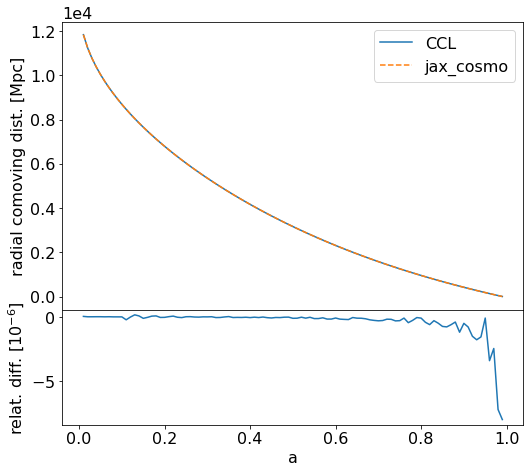}
    \caption{Comparison of the radial comoving distance between CCL and \jaxcosmo. More plots are available in the  companion notebook \nblink{CCL_comparison}.} 
    \label{fig:chi_comparison}
\end{figure}

\begin{figure}
    \centering
    \includegraphics[width=\columnwidth]{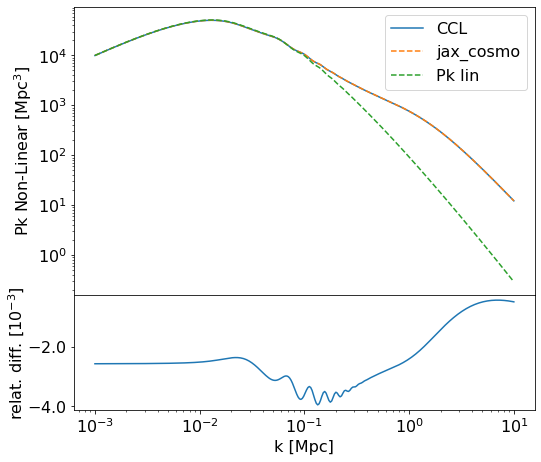}
    \caption{Comparison of the non-linear matter power spectrum (\textit{halofit} function) between CCL and \jaxcosmo. Also shown is the linear power spectrum.}     \label{fig:halofit_comparison}
\end{figure}

\begin{figure}
    \centering
    \includegraphics[width=\columnwidth]{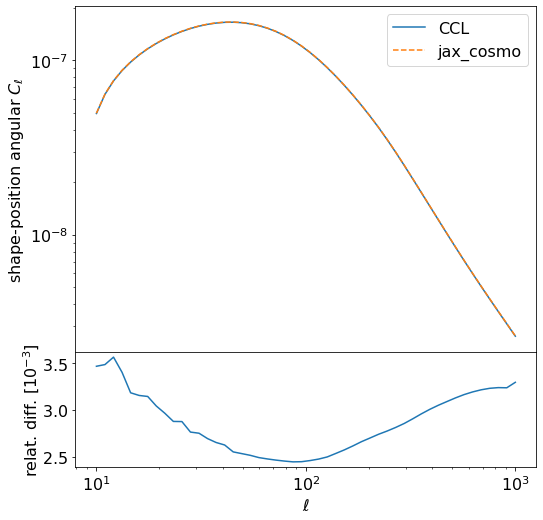}
    \caption{Comparison of the shape-position angular power spectrum between CCL and \jaxcosmo.} 
    \label{fig:Cell_comparison}
\end{figure}

To illustrate the different features available with the present version of the library (\jaxcosmo\ \texttt{0.1}, which is available in the Python Package Index PyPI\footnote{\url{https://pypi.org/}}), we have written a  companion notebook \nblink{CCL_comparison} to compare it to the well-validated  Core Cosmology Library \citep{2019ApJS..242....2C}\footnote{\url{https://ccl.readthedocs.io}, version \texttt{2.5.1}.}. As examples, Figures \ref{fig:chi_comparison},
\ref{fig:halofit_comparison} and \ref{fig:Cell_comparison}
show the radial comoving distance (Eq.~\ref{eq:radial_comoving}), the non-linear matter power spectrum computation, and the angular power spectrum for galaxy-galaxy lensing (Eq.~\ref{eq:Cell_limber}) using the \texttt{NumberCounts} and \texttt{WeakLensing} kernel functions. \jaxcosmo\ features a suite of validation tests against CCL, automatically validating the precision of all computations to within the desired numerical accuracy; the relative differences between the two libraries are at the level of few $10^{-3}$ or better.

These numerical differences are mostly due to different choices of integration methods and accuracy parameters (e.g. number of quadrature points). Increasing these parameters leads to performance degradation for \jaxcosmo\, but increases the XLA compilation memory requirements significantly, especially for the angular power spectra computation. Since these differences are likely to be within the tolerance of the current generation of cosmological surveys, this trade-off is an acceptable one.

\resubnote{Despite this high precision with respect to the CCL library, the current \jaxcosmo\ implementation will have to be reviewed to tackle Stage IV accuracy requirements. For instance the non-limber $C_\ell$  computation \citep{n5k, angpow} would be a limitation needing revision in a future release.}


% \section{Unleashing the Power of Fisher Forecasts with Automatic Differentiation}
\section{Fisher Information Matrices Made Easy}
\label{sec-fisher-forecast}

As a first illustration of the value of \textit{autodiff} in cosmological computations, we present in this section a few direct applications involving the computation of the Fisher information matrix, widely used in cosmology \citep{1997ApJ...480...22T,Stuart1991}. \\
Not only does the computation of the Fisher matrix become trivial, but the Fisher matrix itself becomes differentiable, allowing in particular for powerful survey optimization applications.

%JEC 5 August 22: not sure that should be given here as we have an Appendix for that and the main application described in the following sections concern DES Y1. 
%\subsection{DES Y1 setting}
% Here we describe the data and 3x2pt model for DES Y1 
% In this section, we adopt as a setting a DES Y1 3x2pt-like analysis. 

\subsection{Instantaneous Fisher Forecasts}
Fisher matrices are a key tool in cosmology forecasting and experimental planning. By computing the Hessian matrix of a likelihood with respect to its parameters, we can find a Gaussian approximation to a posterior, which is usually sufficient for comparing the constraining power of different experimental configurations. As noted above, computing Fisher matrices is notoriously error-prone, since finite difference approximations to likelihoods must be carefully tuned for convergence, and observable calculations can be numerically unstable. Autodiff can help avoid this challenge.

We first illustrate, with an artificial case study, the computation of a Fisher matrix using two methods with the \textit{autodiff} ability of \jax. For the detailed implementation, the reader is invited to look at the following companion notebook \nblink{Simple-Fisher}. In this example we use four tracer redshift distributions: two to define \texttt{WeakLensing} kernels and two for \texttt{NumberCounts} kernels. Then, the $10$ angular power spectra $C_\ell^{p,q}$ ($p,q:1,\dots,4$) are computed for $50$  logarithmically-spaced angular moments between $\ell=10$ and $\ell=1000$ using Equation \ref{eq:Cell_limber}. The Gaussian covariance matrix is computed simultaneously. A dataset is obtained from the computation of the $C_\ell^{p,q}$ with a fiducial cosmology. Then, the following snippet shows the log likelihood function $\mathcal{L}(\theta)$ implementation considering a constant covariance matrix ($\theta$ stands for the set of cosmological parameters). 
%\begin{minted}[fontsize=\footnotesize]{python}
\begin{lstlisting}[language=iPython]
@jax.jit
def likelihood(p):
    # Create a new cosmology at these parameters
    cosmo = jc.Planck15(Omega_c=p[0], sigma8=p[1])
    # Compute mean and covariance of angular Cls
    mu, C = jc.angular_cl.gaussian_cl_covariance_and_mean(cosmo, ell, tracers, sparse=True)
    # Return likelihood value assuming constant covariance, so we stop the gradient
    # at the level of the precision matrix, and we will not include the logdet term
    # in the likelihood
    P = jc.sparse.inv(jax.lax.stop_gradient(C))
    r = data - mu
    return -0.5 * r.T @ jc.sparse.sparse_dot_vec(P, r)
\end{lstlisting}
%\end{minted}
The \texttt{jc.sparse} functions are implementations of block matrix computations: a sparse matrix is represented as a 3D array of shape $(n_y, n_x, n_{diag})$ composed of $n_y \times n_x$ square blocks of size $n_{diag} \times n_{diag}$.  The \texttt{jax.jit} decorator builds a compiled version of the function on first use. 

The first approach to obtaining approximate 1-sigma contours of the two parameters ($\Omega_c, \sigma_8$) with a Fisher matrix uses the Hessian of the log-likelihood as follows:
\begin{equation}
F_{i,j} = - \frac{\partial^2\mathcal{L}(\theta)}{\partial \theta_i \partial \theta_j}
\qquad (\theta_1=\Omega_c, \theta_2=\sigma_8)
\label{eq:fisher_way1}
\end{equation}
which is accomplished in two lines of \jax\ code:
\begin{lstlisting}[language=iPython]
hessian_loglike = jax.jit(jax.hessian(likelihood))
F = - hessian_loglike(params)
\end{lstlisting}

The second approach to computing the Fisher matrix, which is restricted to Gaussian likelihoods but is more commonly used in the field because of its numerical stability, is to define a function that computes the summary statistic mean  $\mu(\ell; \theta)$; the Fisher matrix elements are then:
\begin{equation}
    F_{i,j} = \sum_\ell \frac{\partial \mu(\ell)}{\partial \theta_i}^T C^{-1}(\ell)\frac{\partial \mu(\ell)}{\partial \theta_j}
    \label{eq:fisher_way2}
\end{equation}
where $C^{-1}(\ell)$ is the covariance matrix computed with the fiducial cosmology. This can be computed in \jax\ as: 
\begin{lstlisting}[language=iPython]
# We define a parameter dependent mean function
@jax.jit
def jc_mean(p):
    cosmo = jc.Planck15(Omega_c=p[0], sigma8=p[1])
    # Compute signal vector
    mu = jc.angular_cl.angular_cl(cosmo, ell, tracers)
    # We want mu in 1d to match the covariance matrix
    return mu.flatten() 
# We compute its jacobian with JAX, and we JIT it for efficiency
jac_mean = jax.jit(jax.jacfwd(jc_mean))
# We can now evaluate the jacobian at the fiducial cosmology
dmu = jac_mean(params)
# Now we can compose the Fisher matrix
F = jc.sparse.dot(dmu.T, jc.sparse.inv(cov), dmu)
\end{lstlisting}

\begin{figure}
    \centering
    \includegraphics[width=0.7\columnwidth]{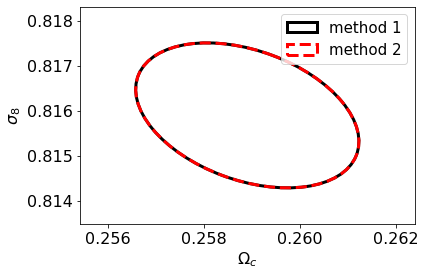}
    \caption{Comparison of the two methods to compute the Fisher matrix: Eq.~\ref{eq:fisher_way1} (method 1) and Eq.~\ref{eq:fisher_way2} (method 2).} 
    \label{fig:simple_fisher_1}
\end{figure}

\begin{figure}
    \centering
    \includegraphics[width=\columnwidth]{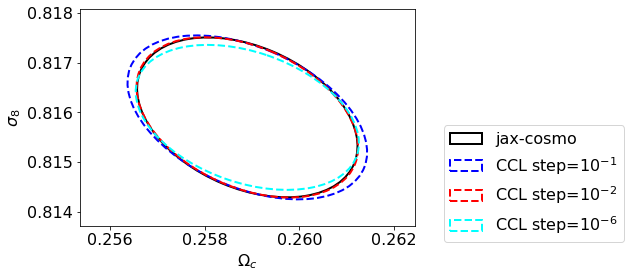}
    \caption{Fisher matrices (Eq.\ref{eq:fisher_way1}) estimated using the CCL for the angular power spectra estimation and finite difference method to get the Jacobian  with different spacing of parameters ($10^{-6}$, $10^{-2}$, $10^{-1}$). For comparison, the contour obtained with \jaxcosmo\ is reproduced from Figure \ref{fig:simple_fisher_1} in black.} 
    \label{fig:simple_fisher_2}
\end{figure}

\jax\ implementations of the two methods agree to near perfect accuracy, as shown in Figure \ref{fig:simple_fisher_1}.
It is worth noting that in the two methods for Fisher matrix computations described above, the user does not need to vary individual parameter values to compute the 1st or 2nd order derivatives; this is in contrast to the usual finite difference methods. As an illustration, we have used CCL to compute the Fisher matrix via Equation \ref{eq:fisher_way2}. To do so, the Jacobian ($\partial \mu/\partial\theta_\alpha$) is computed with centered  order 4 finite differences available in the \texttt{Numdifftools} Python package\footnote{\url{https://numdifftools.readthedocs.io}}. Using the parameter values spaced by ($10^{-6}$, $10^{-2}$, $10^{-1}$) one obtains three different approximation of the 1-sigma contours as shown on Figure \ref{fig:simple_fisher_2}. The contour that agrees best with the \jaxcosmo\ method is obtained with the intermediate spacing parameter $10^{-2}$, implying that the user must tune this parameter carefully. Although very simple, this case study demonstrates the significant challenge of using finite difference methods for computing the Fisher matrix, as has been shown for instance in a more advanced case study in \citet{2021arXiv210100298B}. 

\subsection{Survey Optimization by FoM Maximization}
\label{sec-FOM-optimisation}
% Needs to mention the tomo challenge, mention that we can backpropagate through a NN and the cosmology model
% Maybe this can be only mentioned in the discussion.
% \FrL{Insist here on the fact that the FoM itself is differentiable.}
Fisher forecasts are also commonly used in survey and analysis strategy, where running a full posterior analysis for each possible choice would be unfeasible. The inverse area of a projection of a Fisher matrix in parameters of interest can be used as a metric for survey constraining power, such as in the Dark Energy Task Force report \citep{2006astro.ph..9591A}.

\jaxcosmo\ was used in a recent example of such a process, for the  the LSST-DESC 3x2pt tomography optimization challenge \citep{2021OJAp....4E..13Z}, where the best methodology for assignment of galaxies to tomographic bins was assessed using several such figures of merit and related calculations. The \jaxcosmo\ metric proved to be stable and fast.

Because \jax\ functions are differentiable with respect to all their inputs, including survey configuration parameters (e.g. depth, area, etc), we can even compute the derivative of an FoM with respect to these inputs, allowing for rapid and complete survey optimization.

\subsection{Massive Optimal Compression in 3 Lines}

% \FrL{Here we can implement MOPED for the compression}

Once the Fisher matrix has been accurately estimated, the MOPED\footnote{Massively Optimised Parameter Estimation and Data compression} algorithm can be used to compress data sets with minimal information loss \citep{2000MNRAS.317..965H,2016PhRvD..93h3525Z, 2017MNRAS.472.4244H}. In the case of the constant covariance matrix the algorithm compresses data in a way that is lossless at the Fisher matrix level (i.e. Fisher matrices estimated using the compressed and full data are identical, by construction)  which reduces a possibly large data vector $\mu$ of size $N$ to $M$ numbers, where $M$ is the number of parameters $\theta_i$ considered. For instance, in the previous section, $N=500$ as $\mu=(C_\ell^{p,q})$ and $M=2$ for $(\Omega_c,\sigma_8)$. 

The algorithm computes by iteration $M$ vectors of size $N$ such that (taking the notation of the previous section)

\begin{equation}
    b_i = \frac{C^{-1}\mu_{,i}-\sum_{j=1}^{i-1}(\mu_{,i}^T b_j)b_j}{\sqrt{F_{i,i}-\sum_{j=1}^{i-1}(\mu_{,i}^T b_j)^2}}
\label{eq:moped}
\end{equation}
where $\mu_{,i}=\partial \mu/\partial \theta_i$ ($i=1,\dots,M$). The vectors $(b_i)_{i\leq M}$ satisfy the following orthogonality constraint
\begin{equation}
    b_i^T C b_j = \delta_{i,j}
\end{equation}
The algorithm is similar to the Gram-Schmidt process, using the constant covariance matrix $C$ to define the scalar product $\langle b_i, b_j\rangle$. Then, the original data set $x=C_\ell^{p,q}$ is compressed in a data set composed of $M$ numbers $y_i$ according to
\begin{equation}
    y_i = b_i^T x
\end{equation}
These numbers are uncorrelated and of unit variance and this construction ensures that the log-likelihood of $y_i$ given  $\theta_i$ is identical to that of $x$ up to second order, meaning that the Fisher matrices derived from the two parameters should be identical, and in general the $y$ values should lose very little information compared to the full likelihood.

In problems where a (constant) covariance matrix is estimated from simulations, the number of such simulations required for a given accuracy typically scales with the number of data points used. 
MOPED therefore greatly reduces this number, often by a factor of hundreds. Since the uncertainty in the covariance due to a finite number of simulations must be accounted for \citep{2018MNRAS.473.2355S,2007A&A...464..399H}, this reduction can also offset any loss of information from the compression. Inaccuracies in the full covariance matrix used in the data compression result only in information loss, not bias, as does mis-specification of the fiducial parameter set $\theta_i$ in equation \ref{eq:moped}.

% \FrL{hummm but we still do need a full covariance in the compression algorithm, right?} \FrL{And we still do need to invert it... is the idea rather that if we dont have a perfect matrix at that stage the worse that can happen is sub optimal compression?}
%JEC{Well there are several ways that $C$ can be questioned: parameter dependence (i.e. fiducial model/true model), non-gaussianities, errors... So not sure that all these effects induce only sub-optimal compression. But it may be out of the scope of this paragraph/paper.}

Another key advantage of the MOPED algorithm is to eliminate the need for large covariance matrix inversion of size $N\times N$ requiring $O(N^3)$ operations. This inversion takes place not only for the Fisher matrix computation (Eq.~\ref{eq:fisher_way2}), but more importantly in the log-likelihood computation (see the snippet in the previous section). The MOPED algorithm reduces the complexity to $O(M^3)$ operations.

\begin{figure}
    \centering
    \includegraphics[width=\columnwidth]{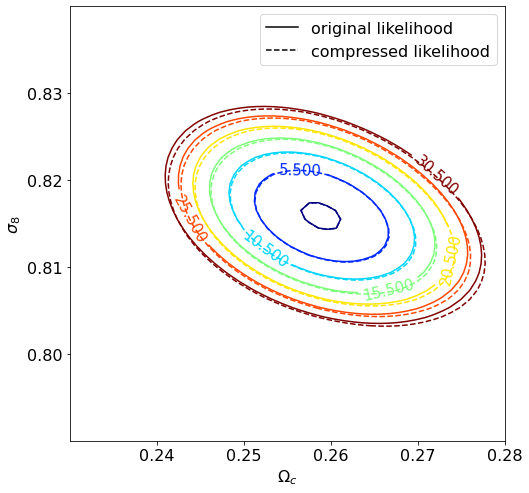}
    \caption{Illustration of the log-likelihood contours obtained with an uncompressed data set (plain lines) and a MOPED version (dashed lines).} 
    \label{fig:moped}
\end{figure}

To give an illustration, the following snippet uses the mock data set and results on the Fisher matrix computation from the previous section, to obtain the MOPED compressed data set composed of two parameters $(y_0,y_1)$ with maximal information on $(\Omega_c, \sigma_8)$:  
\begin{lstlisting}[language=iPython]
# Orthogonal vectors
b0 = jc.sparse.dot(C_inv,dmu[:,0])/jax.numpy.sqrt(F[0,0])
a = dmu[:,1].T @ b0
b1 = (jc.sparse.dot(C_inv,dmu[:,1]) - a * b0)/jax.numpy.sqrt(F[1,1]-a*a)
# MOPED vectors
y0 = b0.T @ data
y1 = b1.T @ data
\end{lstlisting}
Then, the log-likelihood can be implemented as:
\begin{lstlisting}[language=iPython]
@jax.jit
def compressed_likelihood(p):
    # Create a new cosmology at these parameters
    cosmo = jc.Planck15(Omega_c=p[0], sigma8=p[1])
    # Compute mean Cl
    mu = jc.angular_cl.angular_cl(cosmo, ell, tracers).flatten()
    # likelihood using the MOPED vector
    return -0.5 * ((y0 - b0.T @ mu)**2 + (y1 - b1.T @ mu)**2)
\end{lstlisting}
The comparison between contour lines obtained with the original likelihood (uncompressed data set) and the MOPED version are shown in Figure \ref{fig:moped} for the case study of the previous section. Close to the negative likelihood minimum, the lines agree very well.
The case where the covariance matrix depends upon the parameters is discussed by  \citet{2017MNRAS.472.4244H} and leads to similar advantages of the MOPED compression algorithm.

\section{Posterior Inference made fast by Gradient-Based Inference Methods}
\label{sec:chmc}
In the following sections we review more statistical methods which directly benefit from the automatic differentiablity of \jaxcosmo\ likelihoods. We demonstrate a range of gradient-based methods, from Hamiltonian Monte Carlo (HMC), and its \textit{No-U-Turn Sampler} (NUTS) variant, to Stochastic Variational Inference (SVI). We further report their respective computational costs in a DES-Y1 like analysis. All  the methods have been implemented using the \numpyro\ probabilistic programming language (PPL). 
%
% and NUTS after a \textit{Neural Transport} perform using a \textit{Stochastic Variational Inference} (aka SVI). All  the methods have been implemented using the \numpyro\ probabilistic programming language (PPL). 
%
%

\subsection{Description of the DES-Y1 exercise}
\label{sec-DESY1}
\begin{figure}
\centering
\includegraphics[height=5cm]{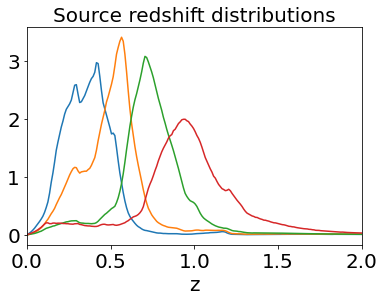}\\
\includegraphics[height=5cm]{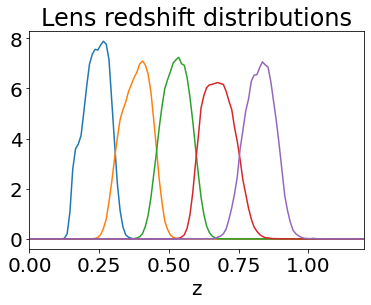}
\caption{Distributions of the sources and lenses for the different redshift bins considered.}
\label{fig-DESY1-src-lens-redshift}
\end{figure}

%JEC{Transfer of the model context from appendix here to ease the reading.}
From the DES Year 1 lensing and clustering data release\footnote{\url{http://desdr-server.ncsa.illinois.edu/despublic/y1a1_files/chains/2pt_NG_mcal_1110.fit}} we have extracted the $N(z)$ distributions of the four source and five lens samples. We normalize the sources to $[1.47, 1.46, 1.50, 0.73]$ effective number of sources per $\mathrm{arcmin}^2$. These distributions are modelled in \jaxcosmo\ using a kernel density estimation in the \texttt{jc.redshift.kde\_nz} function and are presented in Figure \ref{fig-DESY1-src-lens-redshift}.
\begin{table}[htb]
\caption{Priors on the 21 variables of the DES-Y1 of the 3x2pt likelihood (number counts and shear).}
\label{tab-DESY1}
 \centering
\begin{tabular}{ccccccccccc}
\hline
 parameter &  prior \\
 \hline
  \multicolumn{2}{c}{\textbf{Cosmology}} \\
$\Omega_c$ & $\mathcal{U}[0.10, 0.9]$ \\ %JEC 21/10/22 there was exchanges in Omega_c, sigma8, Omega_b intervalle (but ok in the notebooks)
$\sigma_8$ & $\mathcal{U}[0.40, 1.0]$ \\
$\Omega_b$ & $\mathcal{U}[0.03, 0.07]$ \\
$w_0$ & $\mathcal{U}[-2.00, -0.33]$ \\
$h$ & $\mathcal{U}[0.55, 0.91]$ \\
$n_s$ & $\mathcal{U}[0.87, 1.07]$ \\
\multicolumn{2}{c}{\textbf{Intrinsic Alignment}} \\
$A$ & $\mathcal{U}[-5,5]$ \\
$\eta$ &$\mathcal{U}[-5,5]$ \\
\multicolumn{2}{c}{\textbf{Lens Galaxy Bias}} \\
$(b_i)_{i:1,\dots,5}$ & $\mathcal{U}[0.8,3.0]$ \\
\multicolumn{2}{c}{\textbf{Shear Calibration Systematics}} \\
$(m_i)_{i:1,\dots,4}$ & $\mathcal{N}[0.012,0.023]$ \\
\multicolumn{2}{c}{\textbf{Source photo-$z$ shift}} \\
$dz_1$ & $\mathcal{N}[0.001,0.016]$ \\ % JEC note the sign is the one we have used but is different from https://arxiv.org/pdf/1708.01530.pdf
$dz_2$ & $\mathcal{N}[-0.019,0.013]$ \\
$dz_3$ & $\mathcal{N}[0.009,0.011]$ \\
$dz_4$ & $\mathcal{N}[-0.018,0.022]$ \\
\hline
\end{tabular}
\end{table}

Using the \numpyro\ PPL, we then set up a forward model following the DES Y1 Pipeline \citep{2018PhRvD..98d3526A}. Here we show some key elements of the implementation; the details and inference examples may be found in the following notebook \nblink{VectorizedNumPyro_block}. The model parameters and their prior distributions are shown in Table \ref{tab-DESY1}.  For instance the $\Omega_c$ parameter is treated as random variable as follows

\begin{lstlisting}[language=iPython]
Omega_c = numpyro.sample("Omega_c", Uniform(0.1, 0.9))
\end{lstlisting}

Then, we generate mock angular power spectra $C_\ell$ from the auto \& cross correlations of the Number Counts and Weak Lensing probes computed with the \texttt{gaussian\_cl\_covariance\_and\_mean} function in the \texttt{jc.angular\_cl} module. The code reads:
\begin{lstlisting}[language=iPython]
# Define the lensing and number counts probe
probes = [jc.probes.WeakLensing(nzs_s_sys, 
                       ia_bias=b_ia,
                       multiplicative_bias=m),
          jc.probes.NumberCounts(nzs_l, b)]
cl, C = gaussian_cl_covariance_and_mean(cosmo, 
                 ell, probes, 
                 f_sky=0.25, sparse=True)
\end{lstlisting}

with \texttt{cosmo} an instance of the \texttt{jc.Cosmology} setting the cosmological parameters generated with the priors, and \texttt{ell} (i.e. $\ell$) a series of 50 angular modes. After encapsulating the code above with input sampled parameters (using \numpyro\ distribution classes) in a function \texttt{model}, we can generate our mock data (\texttt{cl\_obs}). This comes from this model function evaluated at a fiducial cosmology, with random noise generated by \numpyro:

\begin{lstlisting}[language=iPython]
fiducial_model = numpyro.condition(model,
    {"Omega_c":0.2545, "sigma8":0.801, 
    "h":0.682, "Omega_b":0.0485, "w0":-1.,"n_s":0.971,
     "A":0.5,"eta":0.,
     "m1":0.0,"m2":0.0,"m3":0.0,"m4":0.0,
     "dz1":0.0,"dz2":0.0,"dz3":0.0,"dz4":0.0,
     "b1":1.2,"b2":1.4,"b3":1.6,"b4":1.8,"b5":2.0
      })

with seed(rng_seed=42):
    cl_obs, P, C = fiducial_model()
\end{lstlisting}

% The \textit{inference model} looks very similar to the \textit{forward model} excepts that first we fix the angular power spectra covariance matrix $C$ (and its inverse $P$) and secondly we let \numpyro\ generate the model $C_{\ell}$ from the priors:
% \begin{lstlisting}[language=iPython]
% cl = jc.angular_cl.angular_cl(cosmo, ell, 
%                               probes).flatten()
% return numpyro.sample("cl", MultivariateNormal(cl, 
%                             precision_matrix=P,
%                             covariance_matrix=C))
% \end{lstlisting}

These theoretical $C_{\ell}$ are in turn conditioned on the mock $C_\ell$:

\begin{lstlisting}[language=iPython]
observed_model = numpyro.condition(model, {"cl": cl_obs})
\end{lstlisting}

%Then, as an example the optimisation of the cosmological parameters using for instance the NUTS flavour of MCMC can be conducted as follows
%\begin{minted}[fontsize=\footnotesize]{python}
%\begin{lstlisting}[language=iPython]
%nuts_kernel = numpyro.infer.NUTS(observed_model,
%              step_size=1e-1, 
%              init_strategy=numpyro.infer.init_to_median,
%              dense_mass=True,
%              max_tree_depth=7)
%
%mcmc = numpyro.infer.MCMC(nuts_kernel, 
%                          num_warmup=200, 
%                          num_samples=1000,
%                          num_chains=16,
%                          chain_method='vectorized',
%                          progress_bar=False)
%
%mcmc.run(jax.random.PRNGKey(42))
%\end{lstlisting}
%%\end{minted}
%More details can be found in the repository of this paper to perform a \numpyro\ modeling using \jaxcosmo: a structured NUTS sampling described in Section~\ref{sec-NUTS} \nblink{VectorizedNumPyro_block}, and a SVI optimisation followed by NUTS sampling of Section~\ref{sec-Neural-Reparametrisation} \nblink{DESY_Y1_SVI_and_NeutraReparam}. 
How to use this model to perform inference is described in the sections \ref{sec-NUTS} and \ref{sec-SVI}. %For a complete example of this functionality see the attached notebook: \nblink{VectorizedNumPyro_block}.
\subsection{Vanilla Hamiltonian Monte Carlo}
%
% Show Joe's vanilla HMC results against Cobaya
Hamiltonian Monte Carlo (HMC) is an MCMC-type method particularly suited to drawing
samples from high dimensional parameter spaces.  It was introduced in \citet{1987PhLB..195..216D}
and developed extensively since.  See \citet{betancourt} for a full review; we describe
very basic features here.

HMC samples a space by generating particle trajectories through it, using the log-posterior as the negative potential energy of a particle at each point $q$ in the space. Associated with $q$, we introduce an auxiliary $p$ variable as Hamiltonian momentum such that
\begin{equation}
- \log{\cal P}(q) = V(q) \quad H(q,p) = V(q) + U(p)
\end{equation}
where $U(p)$ is a kinetic energy-like term defined by 
\begin{equation}
U(p) = p^T M^{-1} p
\end{equation}
where $M$ is a mass matrix which should be set to approximate the covariance of the posterior. At each sample, a trajectory is initialized with a random momentum $p$, and then Hamilton's equations are integrated:
\begin{align}
\frac{\mathrm{d}p}{\mathrm{d}t} &= - \frac{\partial V}{\mathrm{d} q} = \frac{\partial \log{\cal P}}{\mathrm{d} q} \\
\frac{\mathrm{d}q}{\mathrm{d}t} &= + \frac{\partial U}{\mathrm{d} p} = M^{-1} p
\end{align}
This is also used to set the scale of the random initial velocities. These differential equations may be integrated numerically, taking $L$ small steps of the \textit{leapfrog} algorithm:
\begin{align}
    p_{n+\frac{1}{2}} &= p_n -\frac{\varepsilon}{2} \frac{\partial V}{\mathrm{d} q}(q_n) \\
    q_{n+1} & = q_n +\varepsilon M^{-1} p_{n+\frac{1}{2}} \\
    p_{n+1} &=  p_{n+\frac{1}{2}} -\frac{\varepsilon}{2} \frac{\partial V}{\mathrm{d} q}(q_{n+1})
\end{align}
where $\varepsilon$ is a step size parameter.

Formally, the set of $n_\mathrm{dim}$ momenta are treated as new parameters, and after 
some number of integration steps a final point in the trajectory is compared to the initial one,
and a Metropolis-Hastings acceptance criterion on the total energy $H(q,p)$ is applied. If the trajectory is perfectly simulated then this acceptance is unity, since energy is conserved; applying it allows
a relaxation of the integration accuracy.

The gradients $\partial \log{\cal P} / \mathrm{d} q$ can be estimated using finite differences,
but this requires at least $2 n_{\mathrm{dim}} + 1$ posterior evaluations per point, greatly slowing it
in high dimension, and as with the Fisher forecasting is highly prone to numerical error. Automatically
calculating the derivative, as in \jaxcosmo, makes it feasible and efficient.

Metropolis-Hastings, and related methods like \texttt{emcee} \citep{goodman-weare,emcee},  suffer as dimensionality increases,  as the region of high probability mass (the \textit{typical set}) becomes a very small fraction of the total parameter space volume. At high enough dimension they become a slow random walk around the space and cannot remain in typical set regions.
The dynamics of HMC allows it to make large jumps that nonetheless stay within the region of high posterior.

The tricky part of HMC is that the \textit{leapfrog} algorithm needs tuning to set the number of steps as well as the step size of integration. The next section examines a solution to this problem: the No-U-Turn HMC version.

\subsection{NUTS}
\label{sec-NUTS}
%
% Show the advantage of using NUTS 
% look at difference in efficiency in terms of how many times we need to call the model.

The No-U-Turn Sampler (\textit{NUTS})  variant of the traditional HMC sampler was introduced in \citet{10.5555/2627435.2638586}. It aims to help finding a new point  $x_{i+1}$ from  the current $x_i$ by finding good and dynamic choices for the leapfrog integration parameters in the root HMC algorithms, the step size $\varepsilon$ and the number of steps $L$.

NUTS iterates the leapfrog algorithm not for a fixed $L$, but until the trajectory starts to ``double back'' and return to previously visited region, at the cost of increasing the number of model evaluations. The user has to set a new  parameter  (\texttt{max\_tree\_depth}) which gives as a power of $2$ the maximum number of model calls at each generation.

Both sampler HMC and NUTS are available in the \numpyro\ library. After the forward model creation for the DES-Y1 3x2pt exercise described in section \ref{sec-DESY1}: \begin{itemize}
    \item we apply a transformation to the cosmological, intrinsic alignment and bias parameters to use a consistent uniform prior $\mathcal{U}[-5,5]$ (Table~\ref{tab-DESY1});
    \item
    we use a structured mass matrix $M$ in a block diagonal form with the blocks as the following sets of parameters $(\Omega_b,\Omega_c,\sigma_8,w_0,h)$ and $(b_i)_{i:1\dots5}$. The remaining parameters have uncorrelated masses. This matrix structure is motivated by the expected degree of parameter correlation as shown for instance in Figure \ref{fig_cobaya_NUTS_SVI_bis}.
\end{itemize}

% As an example the optimisation of the cosmological parameters using for instance the NUTS flavour of MCMC can be conducted as follows with \numpyro:
% \begin{lstlisting}[language=iPython]
% nuts_kernel = numpyro.infer.NUTS(observed_model,
%               step_size=1e-1,
%               init_strategy=numpyro.infer.init_to_median,
%               dense_mass=True,
%               max_tree_depth=7)

% mcmc = numpyro.infer.MCMC(nuts_kernel, 
%                           num_warmup=200, 
%                           num_samples=1000,
%                           num_chains=16,
%                           chain_method='vectorized',
%                           progress_bar=False)

% mcmc.run(jax.random.PRNGKey(42))
% \end{lstlisting}

We ran the NUTS sampler using \texttt{numpyro.infer.NUTS} on the DES Y1 likelihood,  with 16 chains of 1,000 samples each after a warm-up phase consisting of 200 samples, with the \texttt{max\_tree\_depth} set to seven (i.e. 128 steps for each iteration). %\FrL{it should explain that vectorized means 16 chains in parallel on a single GPU.}
Using the ``vectorized'' \texttt{numpyro} option we ran all 16 chains simultaneously on a single GPU, made possible by the \jax\ \textit{vmap} mechanism. If one has several GPU devices available, then the using the \jax\ paralellization mechanism (\textit{pmap}), it is further possible to launch the vectorized sampling across the devices, and get back all the MCMC chains. However, these experiments have all been undertaken on single GPUs, either an NVidia Titan Xp (12GB RAM) on a desktop or an NVidia V100 (32GB RAM) at the IN2P3 Computing Centre\footnote{\url{https://cc.in2p3.fr/en/}}. The elapsed time for these experiments was 20 hours.

\begin{figure*}
\centering
\includegraphics[width=1.5\columnwidth]{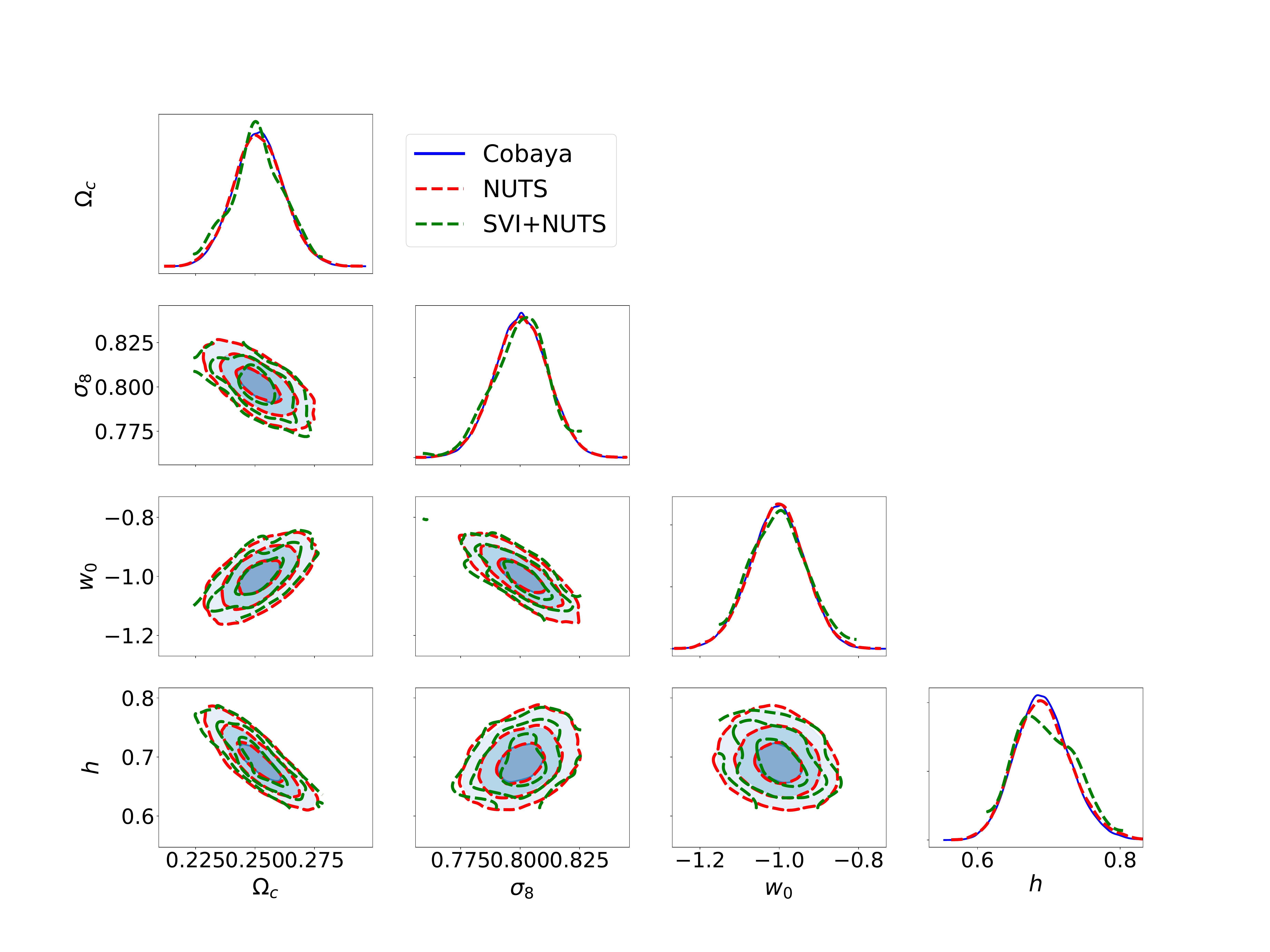}
\caption{Constraints (30\%, 68\%, 90\%) on 4 of the 21 parameters of a simulated DES-Y1 3x2pt likelihood, using the \texttt{Cobaya} Metropolis-Hastings sampler (full curve in blue, \resubnote{140,000} samples), the  NUTS sampling described in section \ref{sec-NUTS} (dashed curve in red; 16,000 samples), and a SVI MVN optimisation followed by a Neural Transport parametrisation to run a NUTS sampling (dashed-dotted curve in green, 200 samples).}
\label{fig_cobaya_NUTS_SVI}
\end{figure*}

The results in terms of relative effective sample sizes (ESS) are detailed in Table~\ref{tab-ESS-NUTS_SVI-1} while the confidence level (CL) contours are presented in Figure \ref{fig_cobaya_NUTS_SVI}. We compare to a reference sample from the highly-optimized \texttt{Cobaya} Metropolis-Hastings implementation \citep{2019ascl.soft10019T,2021JCAP...05..057T}, which is widely used in cosmology and which we ran for around 40 hours on CPU to obtain the set of contours shown.

There is a dramatic improvement of the ESS by about a factor of 10 using the NUTS sampler compared to Cobaya, with very good agreement between the CL contours. It is worth mentioning that the mass matrix structure described above increases the sampling efficiency by about a factor of two.

The speed of the sampling could be further improved: we have tested using the parameter \texttt{max\_tree\_depth=5} and found convergence in five hours, showing a linear scaling in this parameter while keeping the sampling efficiencies at a high level; the user is highly encouraged to tune this critical parameter. 

%\JZ{I don't think the anticipation of the next section was needed here to make the point so I removed it, but please let me know if you're unhappy about this. Otherwise if you're happy please delete this comment.}

% Anticipating the Section~\ref{sec-Neural-Reparametrisation} devoted to the Neural Transport which is an effective solution to validate the model with fewer samples, 
% we have used a NUTS sampling with \texttt{max\_tree\_depth=5} those elapse time is 5 hours. This clearly shows a linear scaling with respect to the number of steps per iteration keeping the sampling efficiencies at a high level (see Table~\ref{tab-ESS-NUTS_SVI-1}), then the user is highly encouraged to tune this critical parameter. 
%

\subsection{Stochastic Variational Inference}
\label{sec-SVI}

We now explore \textit{Stochastic Variational Inference} \citep{10.5555/2567709.2502622, 8588399}, another inference algorithm enabled by auto-differentiation. If we  write $p(z)$ the prior, $p(\mathcal{D}|z)$ the likelihood and $p(\mathcal{D})$ the marginal likelihood, then thanks to Bayes theorem we have $p(z|\mathcal{D})=p(z)p(\mathcal{D}|z)/p(\mathcal{D})$ as the posterior distribution of a model with latent variables $z$ and a set of observations $\mathcal{D}$. Variational Inference (VI) aims to find an approximation to this distribution, i.e. $p(z|\mathcal{D}) \approx q(z;\lambda)$, by determining the variational parameters $\lambda$ of a predefined distribution. To do so, one uses the Kullback-Leibler divergence of the two distributions $KL(q(z;\lambda)||p(z|\mathcal{D}))$ leading to the following relation
\begin{align}
\log p(\mathcal{D}) &= \mathtt{ELBO} +  KL(q(z;\lambda)||p(z|\mathcal{D})) \label{eq-ELBO} \\
\mathrm{with} \ \mathtt{ELBO} &\equiv -\mathbb{E}_{q(z;\lambda)}\left[ \log q(z;\lambda)\right] + \mathbb{E}_{q(z;\lambda)}\left[ \log p(z,\mathcal{D}) \right] 
\end{align}
which defines the \textit{evidence lower bound} (ELBO) that one aims to maximize to get the $\lambda$ values. So, the optimal variational distribution satisfies
\begin{equation}
q(z;\lambda^\ast) = \underset{q(z;\lambda)}{\mathrm{argmax}}\  \mathtt{ELBO} = 
\underset{\lambda}{\mathrm{argmin}}\ \mathcal{L}(\lambda)
\end{equation}
The function $\mathcal{L}(\lambda)$ is the cost function used in practice. It is composed of two parts:
\begin{equation}
\mathcal{L}(\lambda) = \underbrace{\mathbb{E}_{q(z;\lambda)}\left[ \log q(z;\lambda)\right]}_{guide} - \underbrace{\mathbb{E}_{q(z;\lambda)}\left[ \log p(z,\mathcal{D}) \right]}_{model}
\label{eq-loss-svi-1}
\end{equation}
where the \textit{guide} in the \numpyro\ library (i.e. the parameterised function $q$) may be a multi-variate Gaussian distribution (MVN) for instance.

Using the auto-differentiation tool, one can use ``black-box'' guides (aka \textit{automatic differentiation variational inference}). As stated by the authors of \citep{10.5555/3122009.3122023} ADVI specifies a variational family appropriate to the model, computes the corresponding objective
function, takes derivatives, and runs a gradient-based or coordinate-ascent optimization. First we define a invertible differentiable transformation $T$ of the original latent variables $z$ into new variables $\xi$, such $\xi=T(z)$ and $z=T^{-1}(\xi)$, where the new $\xi$ parameters are unbounded, $\xi_i \in (-\infty, \infty)$ and so the subsequent minimization step can be performed with no bound constraints. 
% One can develop ``AutoGuides'' (\numpyro\ terminology) that can be adapted to the user models.
The cost function then reads
\begin{equation}
\mathcal{L}(\lambda) = \underbrace{\mathbb{E}_{q(\xi;\lambda)}\left[ \log q(\xi;\lambda)\right]}_{guide} - \underbrace{\mathbb{E}_{q(\xi;\lambda)}\left[ \log p(\xi,\mathcal{D}) \right]}_{model}
\label{eq-loss-svi-2}
\end{equation}
with
\begin{equation}
p(\xi,\mathcal{D}) \bydef p(T^{-1}(\xi),\mathcal{D}) |J_{T^{-1}}(\xi)|
\end{equation}
which includes the Jacobian of the $T^{-1}$ transformation.
The evaluation of the expectations during gradient descent of the loss Eq.~\ref{eq-loss-svi-2} can be done using what is called \textit{elliptical standardisation} or \textit{re-parametrization trick} or \textit{coordinate transformation} (see references in \citealt{10.5555/2969239.2969303}).  Let us illustrate the method using an invertible transformation  $S_\lambda$ such that $S_\lambda(\xi)=\zeta$, where $\zeta\sim \mathcal{N}(0,I)$\footnote{Notice that $z=T^{-1}(\xi)=(T^{-1} \circ S_\lambda^{-1})(\zeta)=(S_\lambda \circ T)^{-1}(\zeta)=F_\lambda(\zeta)$.}. The Jacobian of this distribution is 1 by definition (volume conservation), so the loss function reads
\begin{multline}
-\mathcal{L}(\lambda) = \underbrace{\mathbb{E}_{\zeta\sim \mathcal{N}(0,I)}\left[ \log p(T^{-1}(S_\lambda^{-1}(\zeta)),\mathcal{D}) + \log |J_{T^{-1}}(S_\lambda^{-1}(\zeta))| \right]}_{model} \\ + \underbrace{\mathbb{H}[q(\xi;\lambda)]}_{guide}
\label{eq-loss-svi-3}
\end{multline}
where $\mathbb{H}(q)\equiv \mathbb{E}_{q(\xi;\lambda)}\left[ \log q(\xi;\lambda)\right]$ is the Shannon entropy of the $q$ distribution; its gradient can be computed once for all for a given $q$ distribution family and reused in any user model.  
Then,
to get $\nabla_\lambda \mathcal{L}$, the $\nabla$ operator can be put inside the expectation which leads to\footnote{To simplify the notation, $T^{-1}(S_\lambda^{-1}(\zeta))$ has been replaced by $z$.}
\begin{multline}
-\nabla_\lambda\mathcal{L}(\lambda) = \mathbb{E}_{\zeta\sim \mathcal{N}(0,I)}\left\{
\left[ \nabla_z \log p(z,\mathcal{D}) \times \nabla_\xi[T^{-1}(\xi)] \right. \right. \\
+ \left. \left. \nabla_\xi \log|J_{T^{-1}}(\xi)| \right] \times \nabla_\lambda S_\lambda^{-1}(\zeta)
\right\}
+ \nabla_\lambda \mathbb{H}[q(\xi;\lambda)]
\label{eq-loss-svi-4}
\end{multline}
%JEC: I think now it is not usefull. A concrete example of the gradient formula is given in Appendix \ref{app-SVI_MVN} with the multivariate Gaussian distribution family used as the guide.

%
%\FrL{here it would be better to have a small statement on the limitations of VI, and why it's a good idea to then go to Neutra.} 
%\FrL{we don't have plots of just the VI?}\JEC{Good point: I have SVI alone with both MVN or Block Neural Autoregressive Flow. See new figure.}
% \JZ{I removed a code snippet here since it was really about numpyro instead of jax-cosmo and felt slightly out of place. Uncomment it if you prefer!}
An implementation example using the \numpyro\ library may be found in this companion notebook: \nblink{DES_Y1_SVI_and_NeutraReparam}.
% Here is the simple way \numpyro\ deals with SVI initialisation and optimisation\footnote{See also listing in Appendix \ref{sec-DESY1}) where the \texttt{model} is described and the \texttt{cl\_obs} are the observed $C_\ell$.}
% \begin{lstlisting}[language=iPython]
% import numpyro.infer.autoguide as autoguide
% from numpyro.infer import SVI, Trace_ELBO
% from numpyro.optim import Adam
% # choose a SVI guide 
% guide = autoguide.AutoMultivariateNormal(model,
%         init_loc_fn=numpyro.infer.init_to_median())
% # SVI optimisation 
% optimizer = numpyro.optim.Adam(1e-3)
% svi = SVI(model_spl, guide,optimizer,loss=Trace_ELBO(num_particles=10))
% svi_result = svi.run(jax.random.PRNGKey(0),20_000, cl_obs)
% #To get samples from SVI approximated posterior 
% samples = guide.sample_posterior(jax.random.PRNGKey(1), svi_result.params, sample_shape=(100_000,))
% \end{lstlisting}
%
Once the optimisation is done one can obtain  i.i.d. $z$ samples from the $q(z,\lambda^\ast)$ distribution applying the inverse of the $S_{\lambda^\ast}$ and $T$ transformations. Using the same DES-Y1 simulation as in previous section, we use both a Multivariate Normal distribution (MVN) and a Block Neural Autoregressive Flow (B-NAF) \citep{pmlr-v115-de-cao20a} as \textit{guides} to approximate the true posterior (\textit{model}). The B-NAF architecture is composed of a single flow using a block autoregressive structure with 2 hidden layers of 8 units each. The SVI optimization has been performed with the Adam optimizer \citep{KingmaB14} and a learning rate set to $10^{-3}$. We have stopped the optimization after 20,000 (30,000) steps to ensure a stable ELBO loss convergence when using the B-NAF (MVN) guides, and no tuning of the optimizer learning rate scheduling or other parameters was performed. This takes about 2 or 3 hours on the NVidia V100 GPU scaling, depending on the number of steps.  

\begin{figure*}
\centering
\includegraphics[width=1.5\columnwidth]{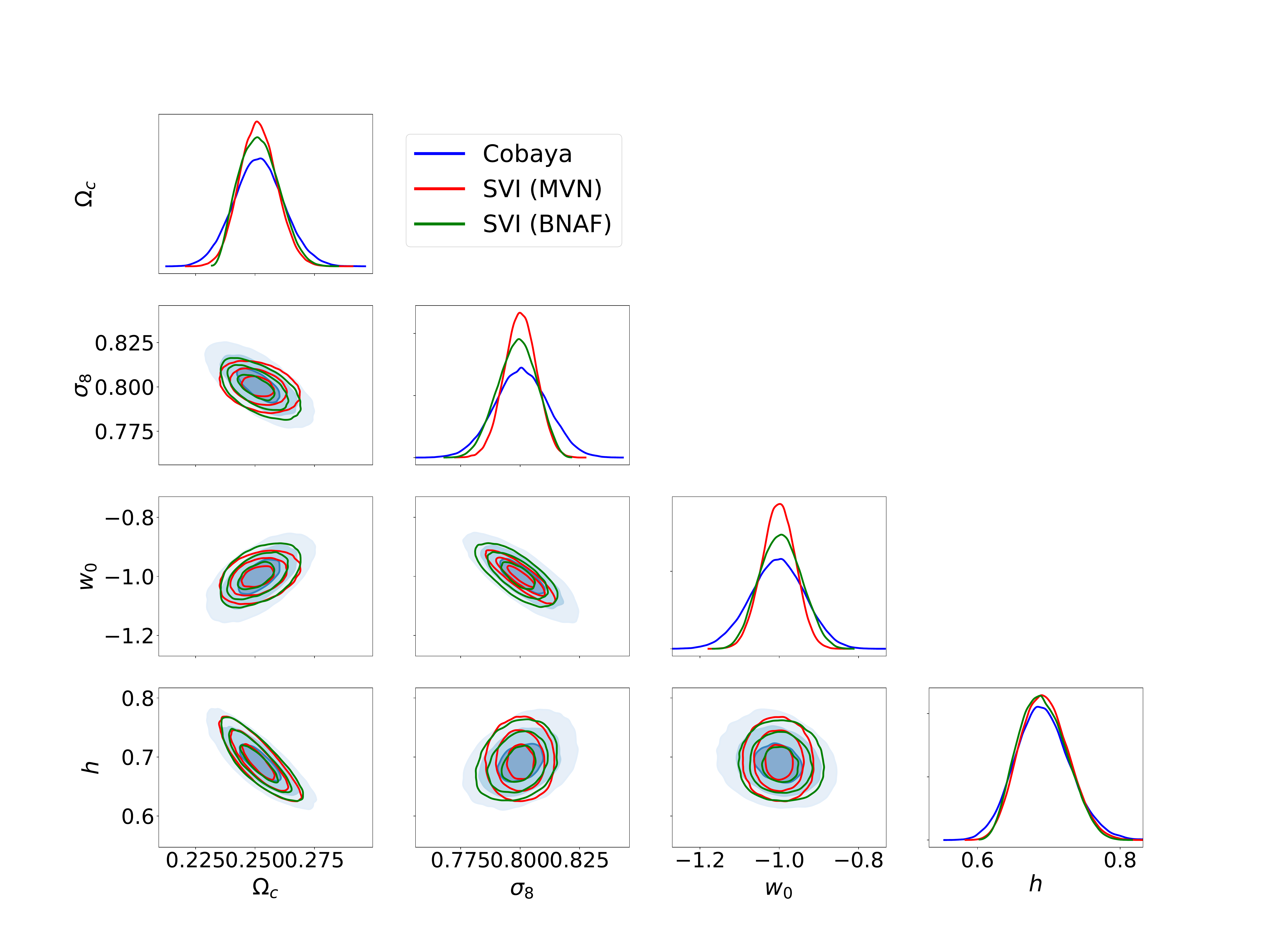}
\caption{Same configuration as in Figure \ref{fig_cobaya_NUTS_SVI} but using sampling of approximated posterior using SVI with MVN and B-NAF guides (see text) compared to \texttt{Cobaya} sampling.}
\label{fig_cobaya_SVIs}
\end{figure*}

In figure \ref{fig_cobaya_SVIs}, we compare the contours obtained with \texttt{Cobaya} (as in figure \ref{fig_cobaya_NUTS_SVI}) and those obtained with the MVN and B-NAF guided SVI.  As noted in \cite{NEURIPS2020_7cac11e2} one challenge with variational inference is assessing how close the variational approximation $q(z,\lambda^\ast)$ is to the true posterior distribution. It is not in the scope of this article to elaborate a statistical diagnosis; rather we show that both guided SVI exhibit rather similar contours and both estimate are close to the Cobaya posterior sampling. The difference is that these SVI approximate posteriors have been obtained in a much shorter time, and can serve as starting point for a NUTS sampler as described in the next section. 
\subsubsection{Neural Transport}
\label{sec-Neural-Reparametrisation}
If the SVI method can be used as is to get $z$ i.i.d. samples from the $q(z,\lambda^\ast)$ distribution as shown on the previous section, the \textit{Neural Transport MCMC} method \citep{Parno2018,2019arXiv190303704H} is an efficient way to boost HMC efficiency, especially in target distribution with unfavourable geometry where for instance the leapfrog integration algorithm has to face squeezed joint distributions for a subset of variables. From SVI, one obtains a first approximation of the target distribution, and this approximation is used to choose a better transform $T$ to map the parameter space to a more convenient one $z=F_\lambda(\zeta)$ (e.g. $F_\lambda=T^{-1}\circ S^{-1}_\lambda$) is such that
\begin{equation}
q(z;\lambda) \rightarrow q(\zeta;\lambda) \bydef q(F_\lambda(\zeta)) |J_{F_\lambda}(\zeta)|
\end{equation}
where $F_\lambda$ with the optimal $\lambda^\ast$ maps the best-fitting $q(z;\lambda^\ast)$ to a geometrically simple function like a unit multivariate normal distribution.  So, one can use a HMC sampler (e.g. NUTS) based on $p(\zeta;\mathcal{D})$ distribution, initialized with $\zeta$ samples from $q(\zeta;\lambda^\ast)$, to get a Markov Chain of $N$ samples $(\zeta_i)_{i<N}$. Then, from the transformation  $z_i=F_{\lambda^\ast}(\zeta_i)$ one finally obtain a Markov Chain with $(z_i)_{i<N}$ samples. % We provide a notebook illustrating this process: \nblink{DES_Y1_SVI_and_NeutraReparam}.
\begin{figure*}
\centering
\includegraphics[width=1.5\columnwidth]{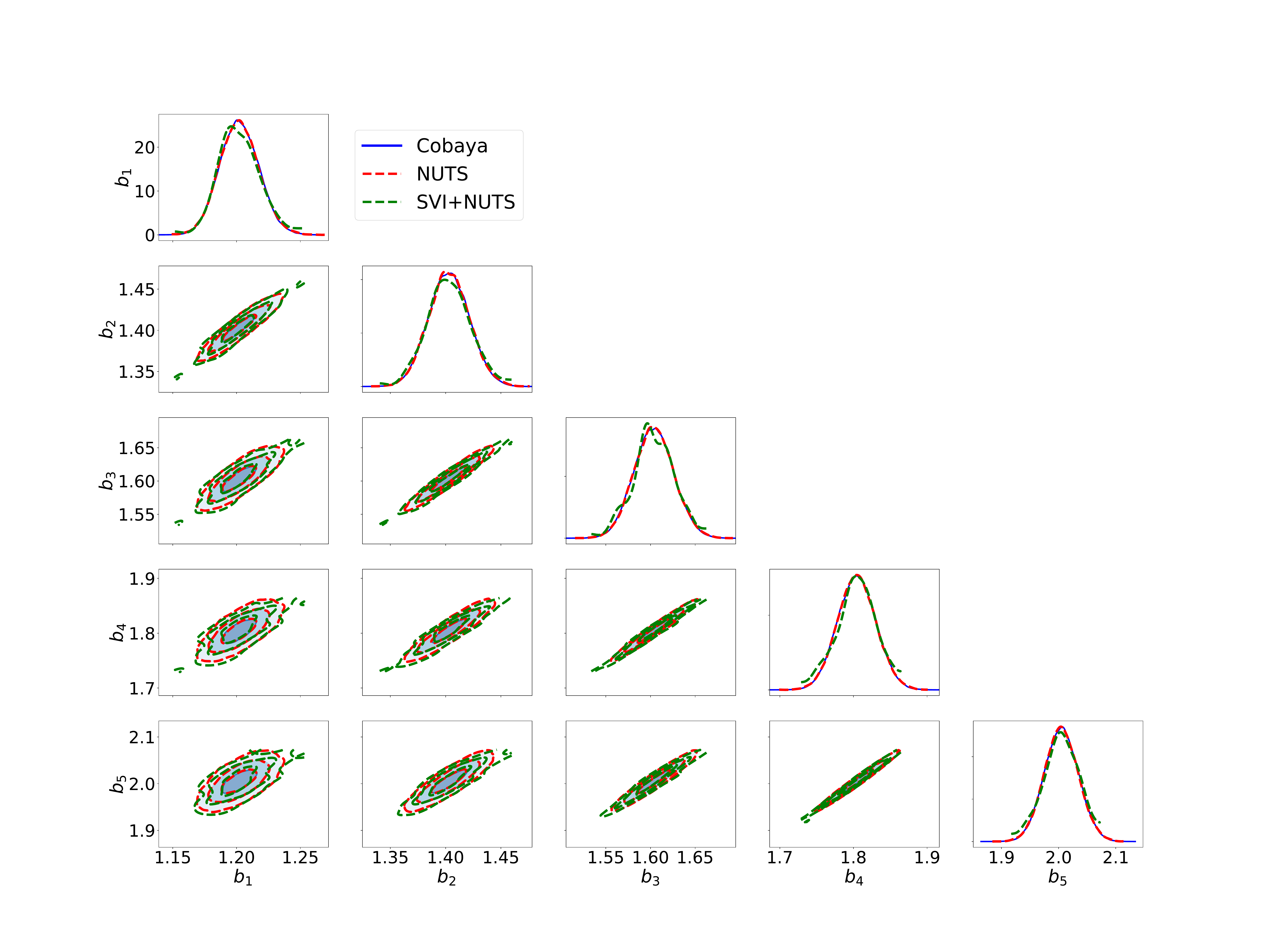}
\caption{Same configuration as in Figure \ref{fig_cobaya_NUTS_SVI} but for the highly correlated five lens galaxy bias.}
\label{fig_cobaya_NUTS_SVI_bis}
\end{figure*}

% \JZ{Deleted code snippet again, but please reinstate if you prefer}.
% The core process is straightforwardly implemented in \numpyro\:
% \begin{lstlisting}[language=iPython]
% from numpyro.infer.reparam import NeuTraReparam
% # Reparametrisation of model to feed NUTS sampler
% neutra = NeuTraReparam(guide, svi_result.params)
% neutra_model = neutra.reparam(model_spl) 
% \end{lstlisting}
%
We have used the MVN guided SVI described in the previous section on the same DES-Y1 analysis described above. NUTS was run using 1 chain of 200 samples (a fast configuration), one chain with 1000 samples and a set of ten chains with 1000 samples each combined into a single run of 10,000 samples (each setup began all chains with 200 samples for initialisation). All NUTS sampling was performed with dense mass matrix optimisation without the special block structuring and with \texttt{max\_tree\_depth=5} which is different than the default NUTS setting described in section \ref{sec-NUTS}. The elapsed time for each of the 3 setups was 50 minutes, 150 minutes and 5 hours, respectively. Naturally, more samples lead to better contour precision. But what is illustrative is the fast configuration results as shown in the Figures \ref{fig_cobaya_NUTS_SVI} and \ref{fig_cobaya_NUTS_SVI_bis}, compared to \texttt{Cobaya} and NUTS results presented in section \ref{sec-NUTS}. The results are good even for the highly correlated lens bias parameters. It is noticeable that running NUTS with the same settings  but without the SVI Neural Transport phase has demonstrated a rather poor behaviour with only 200 samples.

Results in terms of sampling efficiency are shown in Table~\ref{tab-ESS-NUTS_SVI-1}. SVI followed by Neutral Transport gives high efficiency at low number of samples which may be particularly useful during early phases of model development.
\subsection{Sampling efficiency}
\label{sec-results}
\begin{table}[htb]
\caption{The \textit{relative effective sample size} (ESS) in percent computed by the \texttt{Arviz} library \citep{arviz_2019} from: 
%{\color{red} 
%(0) "F.L run" are computed from NUTS sampler with 16 %parallel chains of 1,000 samples each and 200 warm-up %and \texttt{max\_tree\_depth=7} (see text) The numbers %are from \texttt{n\_eff/16,000};
%}
(a): Cobaya \resubnote{140,000} samples;
(b): NUTS sampler with 16 chains of 1,000 samples each and 200 warm-up and \texttt{max\_tree\_depth=7} (Section~\ref{sec-NUTS}); 
(c) SVI Multivariate Normal followed by NUTS and Neural Transform  with one chain of 200 samples and 200 warm-up. The ESS can be larger than 100\% in some cases.  (Section~\ref{sec-Neural-Reparametrisation}).}
\label{tab-ESS-NUTS_SVI-1}
 \centering
\begin{tabular}{ccccccccccc}
\hline
    & $\Omega_b$ & $\Omega_c$ & $\sigma_8$ & $w_0$ & $h$ & $n_s$ & $A$ & $\eta$\\
\hline
(a) &  $3.1$ & $2.5$       & $2.9$      & $2.9$  & $2.6$  & $3.1$  & $3.1$ & $2.8$ \\  
(b) & $48.1$ &  $45.6$     & $36.2$     & $33.4$ & $52.8$ & $50.1$ & $68.8$ & $48.8$\\
(c) & $84.0$ &  $28.0$     & $26.0$     & $20.5$ & $80.0$ & $110.5$ & $58.5$ & $29.5$\\
\hline
\end{tabular}
\end{table}
Looking at the results of Table~\ref{tab-ESS-NUTS_SVI-1}, a key question is: what is the HMC/NUTS gain compared to the highly optimized \texttt{Cobaya} sampler? 
One useful metric is the number of effective samples per model evaluation:
\begin{equation}
    \eta = \frac{n_\mathrm{eff}}{n_\mathrm{eval}} = \frac{N_s \times \varepsilon}{N_s \times n_\mathrm{step}} = \frac{\varepsilon}{n_\mathrm{step}}
\end{equation}
with $N_s$ the total number of samples, $\varepsilon$ the effective sampler efficiency  and $n_\mathrm{step}$ the number of steps (calls) per generated sample. For \texttt{Cobaya} we find $\varepsilon\approx 3\%$ with $n_\mathrm{step}=1$ while for the NUTS sampler $\varepsilon\approx 50\%$ but at the expense of $n_\mathrm{step}=2^5$ or more. With better tuning of the sampling parameters we would expect the $\eta$ values for both methods to become more comparable, but at this intermediate dimensionality the gain from HMC/NUTS compared to a standard MCMC sampler is small. The power of these approaches will become most evident at higher dimensionality still, such as when marginalizing over increasingly complex systematic models. NUTS also makes post-processing simpler, since samples are nearly uncorrelated, removing the need for a \textit{thinning} step, which is a rather delicate procedure, needing know-how to be conducted correctly \citep{doi:10.1146/annurev-statistics-040220-091727, Owen2017}.

% The $\eta$ metric is may be too crude, though, to cover all aspects of the sample generation. \JZ{I don't understand the next point. Doesn't the HMC/NUTS take a long time too on dedicated resources?} One should probably have in mind that the low sampling efficiency of a standard MCMC sampler requires mobilizing a large amount of resources to produce a sufficient large sample batch in a reasonable time scale, i.e. several days on dedicated infrastructure.  %\JZ{I think you need to remove the burn-in when using NUTS too?}.

As the dimensionality of cosmological models increases, methods like HMC/NUTS that by construction are more efficient will become increasingly important. Moreover, using SVI with neural reparametrisation offers an effective way to undertake a progressive validation of a model with rather modest sample set (e.g. starting with 200 samples) producing good enough marginal contours in few hours. In practice, this validation phase can save time before producing sizeable batch for final analysis. The authors have not investigated higher dimensional ($O(10^2)$ parameters) or multi-modal problems, but the key argument in favour of HMC/NUTS sampling is that it exploits the geometry of the typical set of the posterior distribution automatically, unlike the standard random walk of Metropolis-Hasting sampling. Furthermore, using reparametrisation  one can adapt to poor geometry cases (e.g. \citealp{2019arXiv190303704H}).

\section{General discussion}
\label{sec-discussion}
\resubnote{We have demonstrated that the derivatives made available in JAX-Cosmo can be applied in a range of different methodologies to accelerate and improve the stability of cosmological parameter estimation. We expect these improvements to be increasingly important over the coming years, as higher statistical precision pushes us towards more complicated models. Even though the standard cosmological model has now remained static for several decades, models for systematics such as redshift distribution uncertainty and galaxy bias are becoming more and more complex. Using derivative-based methods to constrain them can ensure that we propagate their uncertainties in full.}

\resubnote{With these successes in mind, we now turn to important limitations of \jaxcosmo, and to questions about its place in the cosmology software ecosystem.}

%Lack of autodiff Boltzmann code (emulator)
The first essential barrier to a fully-fledged automatically differentiable cosmology library is the need for a differentiable Boltzmann solver to compute the CMB or matter power spectra. At this stage, \jaxcosmo\ relies on the analytic Eisenstein \& Hu fitting formula for the latter, which is not accurate enough for Stage IV \citep{detf} requirements, and it does not include models beyond $\Lambda$CDM. Existing solvers such as CLASS \citep{2011JCAP...07..034B} or CAMB \citep{camb} are large and complex codes which are not easily reimplemented in an \autodiff\ framework and therefore cannot be directly integrated in \jaxcosmo\ .

A first option to resolve this issue would be to implement from scratch a new Boltzmann code in a framework that supports automatic differentiation. This is the approach behind works such as \texttt{Bolt.jl}\footnote{\url{https://github.com/xzackli/Bolt.jl}} which provides a simplified Boltzmann solver in Julia, or PyCosmo \citep{pycosmo} which is based on the SymPy symbolic mathematics library and could be relatively compatible with \jax. However, even if very promising, both of these options thus far remain limited. While we do believe an automatically differentiable Boltzmann code is the best option, it seems that the cost of developing such a code remains very high at this time.

A second approach would be to develop emulators of a fully-fledged Boltzmann code. Emulators based on neural networks or Gaussian processes are themselves automatically differentiable with respect to cosmological parameters. In fact, the literature is now rich in examples of such emulators \citep[e.g.][and references therein]{bacco1, bacco2, Gunther_2022, nygaard,cosmopower,cosmicnet, emucmb}. After validating their accuracy against a reference CAMB or CLASS implementation, they could be directly integrated as a plug-and-play replacement for the computation of the matter power spectrum. At this time, it seems that using emulators will be the most straightforward approach to bring more accurate models to \jaxcosmo. We believe, though,  that one of the reason for the wide diversity in this is a lack of standardization - a unified interface and validation suite for such methods would provide a much simpler comparison between them and enable wider usage.

\bigskip

% Reasoning behind not making a full emulator of the correlation functions
Connected to this discussion about emulators, a point that could be discussed is whether one even needs a library like \jaxcosmo\ if one can build emulators of a CCL likelihood for use inside gradient-based inference algorithms. While this could indeed be feasible, and of similar cost as just making an emulator of the matter power spectrum, the drawback of this approach is that many analysis parameters and choices become hard coded in the emulator. Since the model for the linear matter power spectrum is typically kept fixed in practical analyses, all the of the choices related in particular to systematics modeling (e.g. photometric redshift errors, galaxy bias, or intrinsic galaxy alignments) will vary significantly in the process of developing the analysis. Building an emulator for the likelihood would require the emulator to be retrained every time the likelihood is changed.

\bigskip

%  - Massively Parallel MCMC on GPU (linked to vmap and/or pmap)
%  - Efficiency of gradient-based inference methods
Another aspect worth discussing are the prospects for scaling and speeding up cosmological inference in practice, given tools such as \jaxcosmo. As we illustrated in the previous section, gradient-based inference techniques yield significantly less correlated MCMC chains, scale better than any other known sampling method as dimension increases, and can provide very fast approximate posteriors if needed. \jaxcosmo\  is also well suited to aid in the parallelisation of likelihoods and algorithms, especially on multiple GPUs, which will become increasingly important as the high-performance computing landscape evolves.

\bigskip

%  - Respective role of CCL and jax-cosmo
Finally, how does \jaxcosmo\ position itself against classical codes such as CCL?  While we are convinced of the benefits of a \jax\ implementation, we expect CCL and other key codes to remain critical as standard cosmology implementations.  Ultimately, a natural transition may occur towards differentiable frameworks like \jaxcosmo\ when they reach the ability to run fully-fledged Stage IV likelihoods.

\section{Conclusions \& Prospects}
\label{sec-conclusion}
%- We have presented a library using automatic differentiation and and automatic GPU offload for a set of key cosmology theory calculations
%- We currently cover only a small number of the many calculations needed in cosmo theory, and welcome contributions for other areas such as CMB.
%- The calculation of Fisher matrices using this method, which is infamous for requiring extensive careful tuning, becomes very simple and robust using JC.
%- We have demonstrated the efficiency in JC of the derivative-aware MCMC methods HMC and NUTS, which are regarded as the only way for samplers to evade the curse of dimensionality up to the hundreds of dimensions we are likely to need for the next generation of surveys.
%- We have shown a proof-of-concept for the use of JC with machine learning methods, opening a whole new space of methods in cosmology.

In this paper, we have presented \jaxcosmo, a cosmology library implemented in the \jax\ framework that enables the automatic computation of the derivatives of cosmological likelihoods with respect to their parameters, and greatly speeds up likelihood evaluations thanks to automatic parallelisation and just-in-time compilation on GPUs. Currently, \jaxcosmo\ contains a small set of features corresponding to a DES-Y1 3x2pt analysis. Being an open source project, contributions of new features for additional scientific areas such as CMB or spectroscopic galaxy clustering are warmly welcome. 

To demonstrate the value of an automatically differentiable library, we have illustrated with concrete examples how Fisher matrices, which are notoriously unstable and require extensive and careful fine tuning, can now be computed robustly and at much lower cost. In addition, Fisher matrices becomes themselves differentiable, which allows for  Figure of Merit optimization by gradient descent, making survey optimization extremely straightforward. 

Going beyond Fisher forecasts, we have also compared simple Metropolis-Hastings to several gradient-based inference techniques (Hamiltonian Monte-Carlo, No-U-Turn-Sampler, and Stochastic Variational Inference). We have shown that the posterior samples with gradient-based methods can reproduce classical methods very efficiently, and can provide approximate posteriors very rapidly. These inference techniques can scale to hundreds of dimensions, and may become necessary in Stage IV analysis, as the number of nuisance parameters is likely to become large.

The next extensions to this framework will be the inclusion of additional cosmological probes, as well as the integration of emulators for the matter power spectrum trained on CAMB or CLASS, as a means to go beyond the current analytic Eisenstein \& Hu model.

In the spirit of reproducible research, all results presented in this paper can be reproduced with code contained in the following GitHub repository:

\url{https://github.com/DifferentiableUniverseInitiative/jax-cosmo-paper/}

%%%
\section*{Credit authorship contribution statement}
\textbf{A. Boucaud:} Software and comments.
\textbf{J.E Campagne:} Conceptualization, Methodology, Software, Validation, Writing, Visualization.
\textbf{S. Casas:} Software contribution for growth rate and power spectra.
\textbf{M.~Karamanis:} Software for Gaussian likelihood computation.
\textbf{D.~Kirkby:} Software and validation for sparse linear algebra.
\textbf{F. Lanusse:} Conceptualization, Methodology, Software, Validation, Writing, Project administration.
\textbf{D. Lanzieri:} Software and validation for redshift distribution.
\textbf{Y. Li:} Software contributions.
\textbf{A. Peel:} Software and validation for spline interpolations in \jax.
\textbf{J. Zuntz:} Investigation, Writing.

%\section*{Declaration of Competing Interest}
%The authors declare that they have no known competing financial interests or personal relationships that could have appeared to influence the work reported in this paper.

%\Acknowledgements
\section*{Acknowledgements}
Some of the numerical experiments have been conducted at the IN2P3 Computing Center (CC-IN2P3 - Lyon/Villeurbanne - France) funded by the Centre National de la Recherche Scientifique.

\bibliographystyle{aa} 
\typeout{}
\bibliography{refs}


\end{document}